\def\VersionLong{}
\def\VersionFinal{}

\ifdefined\VersionLong%
	\newcommand{\LongVersion}[1]{\ifdefined\VersionWithComments{\color{red!40!black}#1}\else#1\fi}
	
\else
	\newcommand{\LongVersion}[1]{\ifdefined\VersionWithComments{\color{black!40}#1}\fi}
	
\fi

\documentclass[runningheads,a4paper]{llncs}

\usepackage[utf8]{inputenc}
\usepackage[english]{babel}

\usepackage[ruled,lined,linesnumbered,noend]{algorithm2e}
\SetKwInOut{Input}{input}
\SetKwInOut{Output}{output}
\SetKw{KwRemove}{remove}
\SetKw{KwAdd}{add}
\SetKw{KwPush}{push}
\SetKw{KwPop}{pop}
\SetKw{KwFrom}{from}
\SetKw{KwTo}{to}
\SetKw{KwCompute}{compute}
\SetKw{KwReturn}{return}
\SetKw{KwBreak}{break}
\SetKw{KwPick}{pick}
\SetKwProg{Fn}{Function}{:}{}
\SetAlgorithmName{Algorithm}{algorithm}{List of Algorithms}

\usepackage{subcaption}
\usepackage{paralist} 
\usepackage{xspace}
\usepackage{versions}

\usepackage{amsmath, amssymb}
\usepackage{mathtools}
\usepackage{stmaryrd}
\usepackage{graphicx}

\ifdefined\VersionWithComments%
	\usepackage{draftwatermark}
	\SetWatermarkText{draft}
	\SetWatermarkScale{15}
	\SetWatermarkColor[gray]{0.9}
\fi

\usepackage[svgnames,table]{xcolor}
\usepackage{colortbl}
\definecolor{darkblue}{rgb}{0.0,0.0,0.6}
\definecolor{darkgreen}{rgb}{0, 0.5, 0}
\definecolor{darkpurple}{rgb}{0.7, 0, 0.7}
\definecolor{darkblue}{rgb}{0, 0, 0.7}

\usepackage[
		colorlinks=true,
	\ifdefined\VersionWithComments{}
		pagebackref=true,
	\fi
		citecolor=darkgreen,
		linkcolor=darkblue,
		urlcolor=darkpurple,
	]{hyperref}
	
\usepackage[capitalise,english,nameinlink]{cleveref} 
\crefname{line}{\text{line}}{\text{lines}} 
\Crefname{line}{\text{Line}}{\text{Lines}} 
\crefname{item}{\text{item}}{\text{items}} 
\crefname{example}{\text{Example}}{\text{Examples}} 
\crefname{assumption}{\text{Assumption}}{\text{Assumptions}} 
\crefname{algorithm}{\text{Algorithm}}{\text{Algorithms}}

\usepackage{wrapfig}
\usepackage{tikz}
\usetikzlibrary{arrows,arrows.meta,automata,positioning}
\tikzstyle{every node}=[initial text=]
\tikzstyle{location}=[rectangle, rounded corners, minimum size=12pt, draw=black, fill=blue!10, inner sep=2pt]
\tikzstyle{final}=[double]
\tikzstyle{accepting}=[final]

\ifdefined\VersionWithComments%
	\usepackage[colorinlistoftodos,textsize=footnotesize]{todonotes}
\else
	\usepackage[disable]{todonotes}
\fi
\newcommand{\gennote}[3]{\todo[linecolor=#2,backgroundcolor=#2!25,bordercolor=#2]{#3: #1}}
\newcommand{\js}[1]{\gennote{#1}{blue}{JS}}
\newcommand{\mw}[1]{\gennote{#1}{orange}{MW}}
\newcommand{\ks}[1]{{\gennote{#1}{purple}{KS}}}
\newcommand{\instructions}[1]{{\gennote{\bfseries #1}{red}{Instructions}}}

\ifdefined\VersionFinal%
\else
	\usepackage[pagewise]{lineno} 
	\linenumbers%
	
\fi

\newcommand{\keywords}[1]{\par\addvspace\baselineskip%
\noindent\keywordname\enspace\ignorespaces#1}

\ifdefined\VersionWithComments%
 	\definecolor{colorok}{RGB}{80,80,150}
\else
	\definecolor{colorok}{RGB}{0,0,0}
\fi

\newcommand{\eg}{\textcolor{colorok}{e.\,g.,}\xspace}
\newcommand{\ie}{\textcolor{colorok}{i.\,e.,}\xspace}

\makeatletter
\def\orcidID#1{\smash{\href{https://orcid.org/#1}{\protect\raisebox{-1.25pt}{\protect\includegraphics{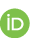}}}}}
\makeatother

\newcommand{\AP}{\mathbf{AP}}
\newcommand{\LTL}{\mathbf{LTL}}

\newcommand{\Next}{\mathcal{X}}
\newcommand{\Until}[1][]{\mathrel{\mathcal{U}_{#1}}}
\newcommand{\Glb}{\Box}
\newcommand{\Evt}{\Diamond}
\newcommand{\Mealy}{\tilde{\mathcal{M}}}
\newcommand{\candidates}{\Psi}
\newcommand{\noIntCandidates}[1][]{\candidates#1_{\mathrm{noInt}}}
\newcommand{\intCandidates}[1][]{\candidates#1_{\mathrm{Int}}}

\begin{document}

\mainmatter%
\title{Efficient Black-Box Checking via Model Checking with Strengthened Specifications
}
\titlerunning{Efficient Black-Box Checking by Specification Strengthening}

%
%
\author{Junya Shijubo\orcidID{0000-0002-2853-1159} \and
 Masaki Waga\orcidID{0000-0001-9360-7490} \and
 Kohei Suenaga\orcidID{0000-0002-7466-8789}}
\authorrunning{}

\institute{
Graduate School of Informatics, Kyoto University, Kyoto, Japan}

%
%

\toctitle{Efficient Black-Box Checking by Specification Strengthening}
\tocauthor{Junya Shijubo et al.}
\maketitle

\thispagestyle{plain}

\ifdefined\VersionWithComments%
	\textcolor{red}{\textbf{This is the version with comments. To disable comments, comment out line~3 in the \LaTeX{} source.}}
\fi

\begin{abstract}
\emph{Black-box checking (BBC)} is a testing method for cyber-physical systems (CPSs) as well as software systems. BBC consists of \emph{active automata learning} and \emph{model checking}; a Mealy machine is learned from the system under test (SUT), and the learned Mealy machine is verified against a specification using model checking. When the Mealy machine violates the specification, the model checker returns an input witnessing the specification violation of the Mealy machine. We use it to refine the Mealy machine or conclude that the SUT violates the specification. Otherwise, we conduct \emph{equivalence testing} to find an input witnessing the difference between the Mealy machine and the SUT.\@ In the BBC for CPSs, equivalence testing tends to be time-consuming due to the time for the system execution. In this paper, we enhance the BBC utilizing model checking with \emph{strengthened specifications}. By model checking with a strengthened specification, we have more chance to obtain an input witnessing the specification violation than model checking with the original specification. The refinement of the Mealy machine with such an input tends to reduce the number of equivalence testing, which improves the efficiency. We conducted experiments with an automotive benchmark. Our experiment results demonstrate the merit of our method.
\end{abstract}


\instructions{Regular Papers (Camera Ready): up to 18 pages (page limits excluding references). No appendix}

\js{hello}
\mw{hello}
\ks{hello}

\section{Introduction}

Due to its safety-critical nature, the safety assurance of a cyber-physical system (CPS) is crucial. However, since a CPS is implemented as a combination of software and physical systems, traditional safety-assurance techniques for software such as testing and formal verification are hard to apply to a CPS.

Much effort has been devoted to adapt these safety-assurance methods for software to a CPS~\cite{DBLP:journals/ngc/Hasuo17}.  Representatives of these methods are \emph{falsification}~\cite{DBLP:conf/rv/FainekosH019} and \emph{formal verification}~\cite{DBLP:conf/dsd/CasagrandeP12,DBLP:conf/se/HerberAL21}.  Given a CPS $\mathcal{M}$ and a specification $\varphi$ that describes how the system should work, a falsification method tries to discover an input to $\mathcal{M}$ that violates $\varphi$ to reveal a flaw of $\mathcal{M}$.  In contrast, a formal verification method tries to guarantee the absence of bugs by mathematically proving that $\mathcal{M}$ conforms to $\varphi$.

There is a tradeoff between these two groups.  Although formal verification ensures high-level safety by resorting to mathematical proofs, its cost is too heavy to be applied to a large CPS.  Furthermore, it cannot be applied if the system $\mathcal{M}$ is a black box.  On the contrary, falsification is cheaper than formal verification and applicable even if $\mathcal{M}$ is a black box.  However, efficiently driving the counterexample search for a black box $\mathcal{M}$ is often challenging.

\emph{Black-box checking (BBC)}~\cite{DBLP:conf/forte/PeledVY99}, one of the falsification methods, is an approach to address this tradeoff.
The main idea of BBC is to combine \emph{active automata learning} such as L*~\cite{DBLP:journals/iandc/Angluin87}, which synthesizes an automaton approximating the behavior of a black-box system, with \emph{model checking}---one of the formal verification techniques---to search for a counterexample in an organized way.

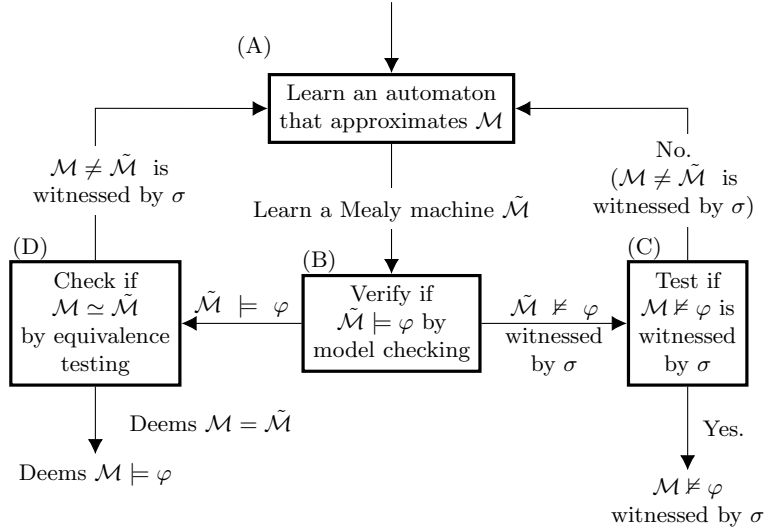
\begin{figure}[t]
  \centering
  \tikzset{
    >={Latex[width=2mm,length=2mm]},
    state/.style={
            rectangle,
            draw=black, very thick,
            minimum height=2em,
            inner sep=4pt,
            align=center,
            },
  }
  \begin{tikzpicture}[node distance=1.6cm,align=center,scale=0.95,every node/.style={transform shape}]
    \node (start) {};
    \node[state, below of=start] (n1)
    {Learn an automaton\\ that approximates $\mathcal{M}$};
    \node[state, below of=n1, yshift=-1.4cm] (n2)
    {Verify if\\$\Mealy \models \varphi$ by\\model checking};
    \node[state, right of=n2, xshift=2.5cm] (n3)
    {Test if\\$\mathcal{M} \nvDash \varphi$ is\\witnessed\\by $\sigma$};
    \node[state, left of=n2, xshift=-2.5cm] (n4)
    {Check if\\$\mathcal{M} \simeq \Mealy$\\ by equivalence\\testing};
    \node[below of=n3, yshift=-0.9cm] (n5) {$\mathcal{M} \nvDash \varphi$ \\witnessed by $\sigma$};
    \node[below of=n4, yshift=-0.5cm] (n6) {Deems $\mathcal{M} \models \varphi$};

    \node[above of=n1, yshift=-0.74cm, xshift=-1.9cm] (n7) {(A)};
    \node[above of=n2, yshift=-0.74cm, xshift=-1cm] (n8) {(B)};
    \node[above of=n3, yshift=-0.55cm, xshift=-0.6cm] (n9) {(C)};
    \node[above of=n4, yshift=-0.55cm, xshift=-0.9cm] (n10) {(D)};

    \draw[->] (start) -- (n1);
    \draw[->] (n1) -- node[text width=4cm,fill=white]
    {Learn a Mealy machine $\Mealy$} (n2);
    \draw[->] (n2) -- node[text width=3cm,yshift=-0.16cm]
    {$\Mealy \nvDash \varphi$ \\witnessed\\by $\sigma$} (n3);
    \draw[->] (n2) -- node[text width=3cm,yshift=0.27cm]
    {$\Mealy \models \varphi$\\} (n4);
    \draw[->] (n3.north) -- ++(0,2.12) -- node[fill=white,xshift=1cm,yshift=-1cm]
    {No.\\($\mathcal{M} \neq \Mealy$ \ is\\witnessed by $\sigma$)} (n1.east);
    \draw[->] (n4.north) -- ++(0,2.12) -- node[fill=white,xshift=-1cm,yshift=-1cm]
    {$\mathcal{M} \neq \Mealy$ \ is\\witnessed by $\sigma$} (n1.west);
    \draw[->] (n3) -- node[xshift=0.5cm] {Yes.} (n5);
    \draw[->] (n4) -- node[xshift=1.6cm]
    {Deems $\mathcal{M} = \Mealy$} (n6);
  \end{tikzpicture}
  \caption{The workflow of black-box checking.}
  \label{fig:bbc-flow}
\end{figure}

\begin{figure}[t]
  \centering
  \tikzset{
    >={Latex[width=2mm,length=2mm]},
    state/.style={
            rectangle,
            draw=black, very thick,
            minimum height=2em,
            inner sep=4pt,
            align=center,
            },
  }
 \begin{tikzpicture}[node distance=1.6cm,align=center,scale=0.89,every node/.style={transform shape}]
    \node (start) {};
    \node[state, below of=start] (n1)
    {Learn an automaton\\ that approximates $\mathcal{M}$};
    \node[state, below of=n1, yshift=-1.4cm] (n2)
    {Verify if\\$\Mealy \models \varphi$ by\\model checking};
    \node[color=red,state, left of=n2, yshift=-0cm,node distance=3.9cm,draw=red] (n2')
    {Verify if\\$\Mealy \models \psi$ by\\model checking};
    \node[color=red,state, below of=n2', node distance=2.5cm,draw=red] (n3')
    {Test if\\ $\mathcal{M} \not\models \psi$ is \\witnessed by $\sigma$};
    \node[state, right of=n2, xshift=2.1cm] (n3)
    {Test if\\$\mathcal{M} \nvDash \varphi$ is\\witnessed\\by $\sigma$};
    \node[state, left of=n2', xshift=-2.1cm] (n4)
    {Check if\\$\mathcal{M} \simeq \Mealy$ by\\ equivalence\\testing};
    \node[below of=n3, yshift=-0.9cm] (n5) {$\mathcal{M} \nvDash \varphi$ \\witnessed by $\sigma$};
    \node[below of=n4, yshift=-1.3cm] (n6) {Deems $\mathcal{M} \models \varphi$};

    \node[above of=n1, yshift=-0.74cm, xshift=-1.9cm] (n7) {(A)};
    \node[above of=n2, yshift=-0.74cm, xshift=-1cm] (n8) {(B)};
    \node[above of=n3, yshift=-0.55cm, xshift=-0.6cm] (n9) {(C)};
    \node[above of=n4, yshift=-0.55cm, xshift=-0.7cm] (n10) {(D)};
    \node[above of=n2', yshift=-0.74cm, xshift=-1cm] (n8') {($\mathrm{B}'$)};
    \node[above of=n3', yshift=-0.6cm, xshift=-0.7cm] (n9') {($\mathrm{C}'$)};

    \draw[->] (start) -- (n1);
    \draw[->] (n1) -- node[text width=4cm,fill=white]
    {Learn a Mealy machine $\Mealy$} (n2);
    \draw[->] (n2) -- node[text width=3cm,yshift=-0.16cm]
    {$\Mealy \nvDash \varphi$ \\witnessed\\by $\sigma$} (n3);
    \draw[->] (n2) -- node[text width=3cm,yshift=0.27cm]
    {$\Mealy \models \varphi$\\} (n2');
    \draw[->,color=red] (n2') -- node[color=red,text width=3cm,yshift=0.27cm]
    {$\Mealy \models \psi$\\} (n4);
    \draw[->,color=red] (n2') -- node[color=red,right]
    {$\Mealy \not\models \psi$ witnessed by $\sigma$} (n3');
    \draw[->,color=red] (n3') -- node[color=red,above right]
    {Yes} (n4);
    \draw[->,color=red] (-2.7,-7.6) -- node[color=red,above,pos=0.35]
    {No. ($\mathcal{M} \neq \Mealy$ \ is witnessed by $\sigma$)} (5.0,-7.6) -- (5.0,-1.2) -- (1.7,-1.2);
    \draw[->] (n3.north) -- ++(0,2.12) -- node[fill=white,xshift=1cm,yshift=-1cm]
    {No.\\($\mathcal{M} \neq \Mealy$ \ is\\witnessed by $\sigma$)} (n1.east);
    \draw[->] (n4.north) -- ++(0,2.12) -- node[fill=white,xshift=-1.6cm,yshift=-1cm]
    {$\mathcal{M} \neq \Mealy$ \ is\\witnessed by $\sigma$} (n1.west);
    \draw[->] (n3) -- node[xshift=0.5cm] {Yes.} (n5);
    \draw[->] (n4) -- node[right,pos=0.7]
    {Deems $\mathcal{M} = \Mealy$} (n6);
 \end{tikzpicture}
  \caption{The workflow of our method, where $\psi$ is a strengthened specification of $\varphi$.  The red part is the changes from the original BBC (\cref{fig:bbc-flow}).}
  \label{fig:our-workflow}
\end{figure}
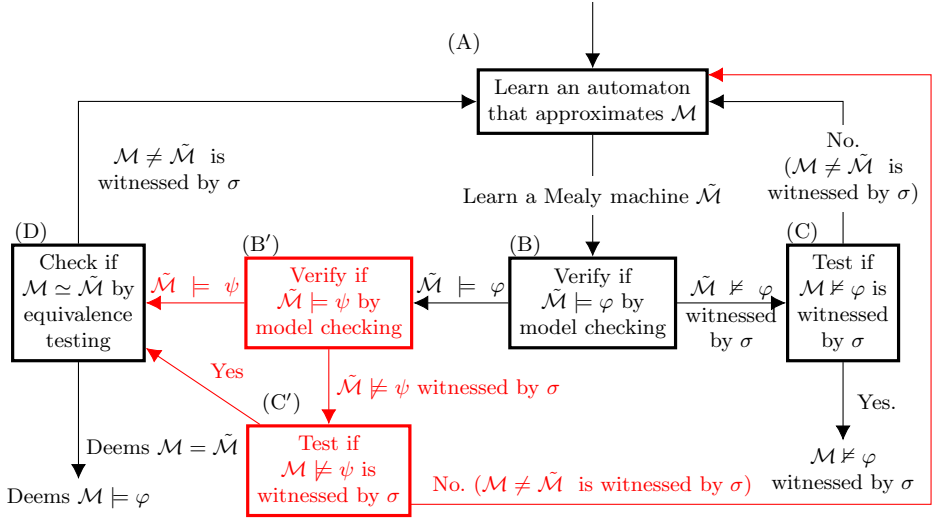

\cref{fig:bbc-flow} shows the workflow of BBC.
It first learns a Mealy machine $\Mealy$ that approximates the behavior of the black-box system $\mathcal{M}$ under test ((A) in \cref{fig:bbc-flow}); this can be done by using the candidate-generation phase of automata learning algorithm such as L*~\cite{DBLP:journals/iandc/Angluin87}.
Notice that the learned $\Mealy$ may not be equivalent to $\mathcal{M}$.
Next, BBC decides whether $\Mealy \models \varphi$ holds by model checking ((B) in \cref{fig:bbc-flow}.)
If this does not hold (i.e., $\Mealy\ \not\models \varphi$), the model-checking procedure returns a counterexample input to $\Mealy$ that drives $\Mealy$ to a state that satisfies $\neg\varphi$.
BBC then checks whether $\sigma$ is a true counterexample or a spurious one by feeding $\sigma$ to the original system $\mathcal{M}$ and observing its behavior ((C) in \cref{fig:bbc-flow}.)
If $\sigma$ is a true counterexample (i.e., $\sigma$ witnesses $\mathcal{M}\nvDash \varphi$), then BBC has disproved $\mathcal{M} \models \varphi$; it returns $\sigma$ as a counterexample.
If $\sigma$ is not a counterexample to the actual system $\mathcal{M}$, then $\sigma$ is a spurious counterexample that exhibits the difference between $\mathcal{M}$ and $\Mealy$.
Then, BBC uses $\sigma$ as a new input to the automata-learning procedure to obtain a new automaton.
If $\Mealy \models \varphi$ holds in the model-checking step in (B), BBC gives $\Mealy$ and $\mathcal{M}$ to an equivalence-testing procedure ((D) in \cref{fig:bbc-flow}).
The equivalence-testing procedure tries to find an input trace that differentiates $\mathcal{M}$ and $\Mealy$ by generating many inputs and executing $\mathcal{M}$ and $\Mealy$.
One may use random sampling for the input generation or may use more sophisticated techniques like hill climbing and evolutionary computation.
If an input $\sigma$ that exhibits the difference between $\mathcal{M}$ and $\Mealy$ is discovered, BBC uses $\sigma$ as a new input to the automata learning procedure.
Otherwise, BBC deems that $\Mealy$ and $\mathcal{M}$ are equivalent and returns $\mathcal{M} \models \varphi$.

One of the practical issues in BBC for CPSs is its long execution time.
In particular, the computational cost of the equivalence testing between a CPS and an automaton is high compared to that of the model checking.
This is because the number of the states of a synthesized automaton to be model-checked is small, but a simulation of the system takes time; therefore, the computational cost of equivalence testing, which requires many runs of simulations, is high.

Based on the above observation, we propose a method to optimize BBC by reducing the number of equivalence tests.
The basic observation is that the number of the equivalence tests conducted by an execution of BBC is the number of the transitions from (B) to (D) in \cref{fig:bbc-flow}; therefore, if we can reduce the number of such transitions, the time spent for an execution of BBC is reduced.

To this end, we adapt BBC so that the model checking of a learned automaton $\Mealy$ is conducted against a \emph{stronger} specification $\psi$ than the original $\varphi$.
A model checking with $\psi$ tends to return a counterexample than it is checked against $\varphi$, which promotes transition from (B) to (C) rather than to (D).

\cref{fig:our-workflow} shows the workflow of the proposed method; the difference from the original BBC is presented in red.
If $\Mealy \models \varphi$ is successfully verified by a model checker ((B) in \cref{fig:our-workflow}), our procedure generates a stronger specification $\psi$ and applies a model checker to verify $\Mealy \models \psi$ (($\mathrm{B'}$) in \cref{fig:our-workflow}).
If the verification fails with a counterexample $\sigma$, our procedure checks whether $\sigma$ witnesses that the original $\mathcal{M}$ violates the strengthened specification $\psi$ (($\mathrm{C'}$) in \cref{fig:our-workflow}).
If it is not the case, $\sigma$ exhibits the difference between $\mathcal{M}$ and $\Mealy$ since $\sigma$ does not drive $\mathcal{M}$ to the violation of $\psi$ but it does for $\Mealy$.
Then, the learned automaton $\Mealy$ is refined by using the new data $\sigma$ ((A) in \cref{fig:our-workflow}).
If $\Mealy$ is verified to conform to $\psi$ or $\sigma$ drives $\mathcal{M}$ to the violation of $\psi$, then our procedure conducts an equivalence test ((D) in \cref{fig:our-workflow}).

To generate a stronger specification $\psi$ than $\varphi$, we define syntactic rewriting rules to strengthen $\varphi$.
The rules include, for example, rewriting of $p \lor q$ to $p \land q$, where $p$ and $q$ are atomic propositions,  and rewriting of an STL formula $\Evt_{I} \varphi$ to $\Evt_{I'} \varphi$, where the interval $I'$ is a subset of $I$.
We define the strengthening relation and prove its correctness.

We implemented our method as an extension of FalCAuN~\cite{DBLP:conf/hybrid/Waga20} that implements BBC for CPSs.
To check the effectiveness of our method, we evaluated our implementation using the Simulink model of an automatic transmission system~\cite{DBLP:conf/cpsweek/HoxhaAF14}.
The result shows that our method is up to 66\% faster than the original BBC, which demonstrates the effectiveness of our method.

\subsection{Related work}

Active automata learning has various applications in software engineering~\cite{DBLP:conf/dagstuhl/HowarS16,DBLP:conf/sfm/SteffenHM11},
\eg{} specification mining~\cite{DBLP:conf/apn/EsparzaLS10,DBLP:conf/sigsoft/2015} and synthesis~\cite{DBLP:conf/fm/LinH14}.
\emph{Black-box checking (BBC)}~\cite{DBLP:conf/forte/PeledVY99}, which is also known as \emph{learning-based testing (LBT)}, is an application of active automata learning for system testing.
BBC has been used for testing numerical software~\cite{DBLP:conf/pts/MeinkeN10}, distributed systems~\cite{DBLP:conf/sefm/MeinkeN15}, and autonomous systems~\cite{DBLP:conf/kbse/KhosrowjerdiM18}.
BBC is implemented in LBTest~\cite{DBLP:conf/icst/MeinkeS13} and LearnLib~\cite{DBLP:conf/cav/IsbernerHS15,DBLP:journals/isse/MeijerP19}.

As one of the quality assurance methods of CPSs,
falsification~\cite{DBLP:conf/rv/FainekosH019,DBLP:series/lncs/BartocciDDFMNS18} has been attracting attention from both academia and industry.
There are several practical tools for falsification, for example, S-TaLiRo~\cite{DBLP:conf/tacas/AnnpureddyLFS11} and Breach~\cite{DBLP:conf/cav/Donze10}. 
See also the report~\cite{ARCH20:ARCH_COMP_2020_Category_Report} of the annual friendly competition on the falsification problem.
There are various industrial case studies utilizing these tools for falsification.
Yamaguchi et al.~\cite{DBLP:conf/fmcad/YamaguchiKDS16} presents a case study that uses the falsification tool Breach to find issues in automotive systems. Hoxha et al.~\cite{DBLP:conf/cpsweek/HoxhaAF14a} demonstrates falsification on industrial size engine model using S-TaLiRo. Cameron et al.~\cite{DBLP:conf/rv/CameronFMS15} uses S-TaLiRo to search for violations of artificial pancreas controllers that automate insulin delivery to patients with type-1 diabetes.

\emph{Robustness-guided falsification}~\cite{DBLP:conf/rv/FainekosH019} is a widely-used technique to solve the falsification problem with optimization, \eg{} simulated annealing~\cite{kirkpatrick1983optimization} and CMA-ES~\cite{DBLP:conf/cec/AugerH05}.\@
Robustness-guided falsification reduces the falsification problem to minimizing the quantitative satisfaction degree called \emph{robustness}~\cite{DBLP:journals/tcs/FainekosP09,DBLP:conf/formats/DonzeM10} of the specification $\varphi$ in \emph{signal temporal logic (STL)}~\cite{DBLP:conf/formats/MalerN04}.
Recently, BBC is also used for the falsification of CPSs~\cite{DBLP:conf/hybrid/Waga20}.
In~\cite{DBLP:conf/hybrid/Waga20}, an equivalence testing dedicated to CPS falsification called \emph{robustness-guided equivalence testing} is introduced.
Robustness-guided equivalence testing tries to find a witness $\sigma$ of $\Mealy \neq \mathcal{M}$ useful for the falsification problem by minimizing the robustness.

\emph{Robust linear temporal logic (rLTL)}~\cite{DBLP:conf/csl/TabuadaN16} is an extension of LTL with 5-valued semantics.
rLTL is used to guarantee that a requirement violation due to a \emph{small} assumptions violation is \emph{small}.
The 5-valued semantics of rLTL is based on a \emph{weakening} of temporal operators in rLTL formulas related to our \emph{strengthening}.

After recalling the preliminaries in \cref{sec:preliminary}, we introduce our enhancement of BBC via model checking with strengthened specifications in \cref{section:our_method}.
We show the experimental evaluation in \cref{section:experiment}, and
conclude in \cref{section:conclusions_and_future_work}.

\section{Preliminaries}\label{sec:preliminary}

For a set $S$, we denote its power set by $\mathcal{P}(S)$.
For a set $S$, an infinite sequence $s = s_0, s_1, \dots \in S^\omega$ of $S$,
and $i, j \in \mathbb{N}, i \leq j$, we denote the subsequence $s_i, s_{i+1}, \dots , s_j \in S^*$ by $s[i, j]$.
For a set $S$, a finite sequence $s \in S^*$ of $S$, and an infinite sequence $s' \in S^\omega$ of $S$, we denote their concatenation by $s \cdot s'$.

\subsection{Linear temporal logic}

\emph{Linear temporal logic (LTL)}~\cite{DBLP:conf/focs/Pnueli77} is a temporal logic which is
commonly used to describe temporal behaviors of systems.

\begin{definition}
  [Syntax of linear temporal logic]
 \label{definition:LTL}
  For a finite set $\AP$ of atomic propositions, the syntax of \emph{linear temporal logic} is defined as follows,
 where $p \in \AP$ and $i, j \in \mathbb{N} \cup \{ \infty \} $ satisfying $i \leq j$\footnote{In the standard definition of LTL,\@ the interval $\Until[[i, j)]$ is always $[0, \infty)$ and it is omitted. We employ the current syntax to emphasize the similarity to STL.\@ We note that this does not change the expressive power.}.
  \[
  \varphi , \psi \Coloneqq \top \mid p \mid \neg \varphi \mid \varphi \vee \psi \mid \varphi \Until[[i, j)] \psi \mid \Next \varphi
  \]
  We denote the set of linear temporal logic formulas by $\LTL$.
\end{definition}

In addition to the syntax in \cref{definition:LTL},
we use the following syntactic abbreviations of LTL formulas. 
Intuitively, $\Evt \varphi$ stands for ``eventually $\varphi$ holds'' and $\Glb \varphi$ stands for ``globally $\varphi$ holds''.
\begin{align*}
& \bot \equiv \neg \top, \quad
  \varphi \wedge \psi \equiv \neg ((\neg \varphi) \vee (\neg \psi)), \quad
  \varphi \rightarrow \psi \equiv (\neg \varphi) \vee \psi, \quad \\
& \Evt_{[i, j)}\varphi \equiv \top \Until[[i,j)] \varphi, \quad
  \Glb_{[i, j)} \varphi \equiv \neg (\Evt_{[i, j)} \neg \varphi), \quad
  \varphi \Until \psi \equiv \varphi \Until[[0, \infty)] \psi \\
& \Evt \varphi \equiv \Evt_{[0, \infty)} \varphi, \quad
  \Glb \varphi \equiv \Glb_{[0, \infty)} \varphi
\end{align*}

The semantics of LTL formulas is defined by the following satisfaction relation $(\pi, k) \models \varphi$.
For an infinite sequence $\pi$, an index $k$, and an LTL formula $\varphi$, $(\pi, k) \models \varphi$ intuitively stands for ``$\pi$ satisfies $\varphi$ at $k$''.

\begin{definition}
  [Semantics of linear temporal logic]
  \label{definition:LTLsemantics}
  For an LTL formula $\varphi$, an infinite sequence $\pi = \pi_0, \pi_1, \dots \in (\mathcal{P}(\AP))^\omega$ of subsets of atomic propositions, and $k \in \mathbb{N}$,
  we define the satisfaction relation $(\pi, k) \models \varphi$ as follows.
  \begin{equation*}
    \begin{alignedat}{3}
    (\pi, k) & \models \top\\
    (\pi, k) & \models p & & \iff & & p \in \pi_k \\
    (\pi, k) & \models \neg \varphi & & \iff & & (\pi, k) \nvDash \varphi \\
    (\pi, k) & \models \varphi \vee \psi & & \iff & & (\pi, k) \models \varphi \vee (\pi, k) \models \psi \\
    (\pi, k) & \models \Next \varphi & & \iff & & (\pi, k + 1) \models \varphi \\
    (\pi, k) & \models \varphi \Until[[i, j)] \psi & & \iff & & \exists l \in [k + i, k + j).\, (\pi, l) \models \psi \\
      & & & & & \wedge \forall m \in \{k, k + 1, \dots, l\}.\, (\pi, m) \models \varphi
    \end{alignedat}
  \end{equation*}
  If we have $(\pi, 0) \models \varphi$, we denote $\pi \models \varphi$.
\end{definition}

In this paper, we mainly use a subclass of LTL called \emph{safety} LTL.\@
Safety LTL is a subclass of LTL whose violation can be witnessed by a \emph{finite} sequence.
The existence of finite witness simplifies the application to BBC.

\begin{definition}
 [safety LTL]\label{def:safety}
 An LTL formula $\varphi$ is \emph{safety} if for any infinite sequence $\pi \in (\mathcal{P}(\AP))^{\omega}$ satisfying $\pi \nvDash \varphi$,
 there is $i \in \mathbb{N}$ such that for any prefix $\pi[0, j]$ of $\pi$ longer than $i$ (\ie{} $j > i$), and
 for any infinite sequence $\pi' \in (\mathcal{P}(\AP))^\omega$, we have $\pi[0, j]\cdot \pi' \nvDash \varphi$
\end{definition}

\subsection{LTL model checking}

Model checking is a technique to verify the correctness of a system model $\mathcal{M}$ against a specification $\varphi$.
We utilize Mealy machines for system modeling and LTL formulas for a specification $\varphi$.

\begin{definition}
 [Mealy machine]
 For an input alphabet $\Sigma$ and an output alphabet $\Gamma$, 
 a \emph{Mealy machine} is a 3-tuple $\mathcal{M} = (L, l_0, \Delta)$,
 where $L$ is the finite set of locations, $l_0 \in L$ is the initial location, and $\Delta : (L \times \Sigma) \to (L \times \Gamma)$ is the transition function.
\end{definition}

For a Mealy machine $\mathcal{M} = (L, l_0, \Delta)$ over $\Sigma$ and $\Gamma$,
the language $\mathcal{L}(\mathcal{M}) \subseteq (\Sigma \times \Gamma)^\omega$ is defined as follows.
\[
  \mathcal{L}(\mathcal{M}) = \{ (a_0, b_0),(a_1, b_1), \dots \mid \exists l_1,l_2, \dots , \forall i \in \mathbb{N}.\,  \Delta(l_i, a_i) = (l_{i+1}, b_i) \}
\]
For an infinite sequence $\sigma = (a_0, b_0),(a_1, b_1), \dots \in (\Sigma \times \Gamma)^\omega$,
we define $\mathbf{pr_1}(\sigma) = a_0, a_1, \dots \in \Sigma^\omega$ and
$\mathbf{pr_2}(\sigma) = b_0, b_1, \dots \in \Gamma^\omega$.
For a Mealy machine $\mathcal{M}$, 
the input language $\mathcal{L}_{in}(\mathcal{M}) \subseteq \Sigma^\omega$ and 
the output language $\mathcal{L}_{\mathit{out}}(\mathcal{M}) \subseteq \Gamma^\omega$ are
$\mathcal{L}_{\mathit{in}}(\mathcal{M}) =
\{ \mathbf{pr_1}(\sigma) \mid \exists \sigma \in \mathcal{L(M)} \}$ and
$\mathcal{L}_{\mathit{out}}(\mathcal{M}) =
\{ \mathbf{pr_2}(\sigma) \mid \exists \sigma \in \mathcal{L(M)} \}$.

In the model checking, we use a Mealy machine $\mathcal{M}$ with the output alphabet $\Gamma = \mathcal{P}(\AP)$ to model the system, and check if all the sequences in its language $\mathcal{L}(\mathcal{M})$ satisfy the LTL formula $\varphi$.
Moreover, if there is a sequence in the language $\mathcal{L}(\mathcal{M})$ and violating the LTL formula $\varphi$, the model checker returns a sequence witnessing the violation.
The formal definition of model checking is as follows.

\begin{definition}
  [LTL model checking]
  Let $\Sigma$ be the input alphabet and let $\AP$ be the set of the atomic propositions.
  Given an LTL formula $\varphi$ over $\AP$ and a Mealy machine $\mathcal{M}$ over $\Sigma$ and $\mathcal{P}(\AP)$,
  \emph{LTL model checking} decides if for any $\pi \in \mathcal{L}_{out}(\mathcal{M})$, we have $\pi \models \varphi$.
  If there is $\sigma \in \mathcal{L(M)}$ satisfying $\mathbf{pr_2}(\sigma) \nvDash \varphi$, the LTL model checker returns such $\sigma$.
 We denote $\forall \pi \in \mathcal{L}_{out}(\mathcal{M}).\, \pi \models \varphi$ by $\mathcal{M} \models \varphi$.
\end{definition}


In this paper, we utilize \emph{safety} LTL formulas in \cref{def:safety}.
For any safety LTL formula $\varphi$ with $\mathcal{M} \nvDash \varphi$,
there is a finite sequence $\sigma \in {(\Sigma \times \mathcal{P}(\AP))}^*$ such that
for any $\sigma' \in {(\Sigma \times \mathcal{P}(\AP))}^{\omega}$ satisfying $\sigma \cdot \sigma' \in \mathcal{L(M)}$,
we have $\mathbf{pr_2}(\sigma \cdot \sigma') \not\models \varphi$.
We use such a finite sequence $\sigma$ as a witness of $\mathcal{M} \nvDash \varphi$.
For the discussion on such a finite witness,
we define the \emph{finite} language $\mathcal{L}^{fin}(\mathcal{M})$ of a Mealy machine $\mathcal{M}$ as
$\mathcal{L}^{fin}(\mathcal{M}) = \{\sigma \in (\Sigma \times \mathcal{P}(\AP))^* \mid \exists \sigma' \in (\Sigma \times \mathcal{P}(\AP))^\omega.\, \sigma \cdot \sigma' \in \mathcal{L(M)}\}$.

\subsection{Signal temporal logic}
\emph{Signal temporal logic (STL)}~\cite{DBLP:conf/formats/MalerN04} is a variant of LTL dedicated to representing behaviors of real-valued signals.
Although the standard definition is for \emph{continuous}-time signals, 
we employ \emph{discrete}-time STL~\cite{DBLP:journals/tcs/FainekosP09} since we use STL for BBC.

\begin{definition}
  [signal]
  For a finite set $Y$ of variables, a (discrete-time) \emph{signal} $\sigma \in (\mathbb{R}^Y)^\infty$ is a finite or infinite sequence of valuations $u_i : Y \to \mathbb{R}$.
  For a finite signal $\sigma = u_0, u_1, \dots , u_{n - 1} \in (\mathbb{R}^Y)^*$, we denote the length $n$ of $\sigma$ by $|\sigma|$.
\end{definition}

\begin{definition}
  [discrete-time STL]\label{def:stl}
  For a finite set $Y$ of variables, the syntax of STL is defined as follows,
  where $y \in Y$, ${\bowtie} \in \{ <, > \}$, $c \in \mathbb{R}$, and $i, j \in \mathbb{N} \cup \{\infty\}$.
  \[
  \varphi, \psi \Coloneqq \top \mid y \bowtie c \mid \neg \varphi \mid \varphi \vee \psi \mid \varphi \Until[[i, j)] \psi \mid \Next \varphi
  \]
\end{definition}

Similarly to LTL, we use the following syntactic abbreviations.
\begin{align*}
  & \bot \equiv \neg \top, \quad
    y \geq c \equiv \neg (y < c), \quad
    y \leq c \equiv \neg (y > c), \quad
    \varphi \wedge \psi \equiv \neg ((\neg \varphi) \vee (\neg \psi)), \\
  & \varphi \rightarrow \psi \equiv (\neg \varphi) \vee \psi, \quad
    \Evt_{[i, j)}\varphi \equiv \top \Until[[i,j)] \varphi, \quad
    \Glb_{[i, j)} \varphi \equiv \neg (\Evt_{[i, j)} \neg \varphi), \\
  & \varphi \Until \psi \equiv \varphi \Until[[0, \infty)] \ \psi, \quad
    \Evt \varphi \equiv \Evt_{[0, \infty)} \varphi, \quad
    \Glb \varphi \equiv \Glb_{[0, \infty)} \varphi
\end{align*}

The semantics of STL formulas is defined similarly to that of LTL formulas.
While the satisfaction of an LTL formula is defined for an infinite sequence $\pi \in (\mathcal{P}(\AP))^\omega$ of a set of atomic propositions, 
the satisfaction of an STL formula is defined for an infinite signal $\sigma \in (\mathbb{R}^Y)^\infty$.
Each inequality constraint in an STL formula is evaluated with the valuation $u_i$ in the signal $\sigma$, and the satisfaction of the other formulas is defined inductively.
Formally, the satisfaction relation $(\sigma, k) \models \varphi$ is inductively defined as follows, where
$\varphi$ is an STL formula over $Y$, $\sigma \in {(\mathbb{R}^Y)}^\omega$ is an infinite length signal over $Y$, and $k \in \mathbb{N}$ is an index.

\begin{equation*}
  \begin{alignedat}{3}
    (\sigma, k) & \models \top \\
    (\sigma, k) & \models y > c & & \iff & & u_k(y) > c \\
    (\sigma, k) & \models y < c & & \iff & & u_k(y) < c \\
    (\sigma, k) & \models \neg \varphi & & \iff & & (\sigma, k) \nvDash \varphi \\
    (\sigma, k) & \models \varphi \vee \psi & & \iff & & (\sigma, k) \models \varphi \vee (\sigma, k) \models \psi \\
    (\sigma, k) & \models \Next \varphi & & \iff & & (\sigma, k + 1) \models \varphi \\
    (\sigma, k) & \models \varphi \Until[[i, j)] \psi & & \iff & & \exists l \in [k+i, k+j). \ (\sigma, l) \models \psi \\
    & & & & & \wedge \forall m \in \{k, k+1, \dots , l \}. \ (\sigma, m) \models \varphi \\
  \end{alignedat}
\end{equation*}

The notion of \emph{safety} is defined similarly to that of LTL.\@
Moreover, model checking with an STL formula is defined similarly.
The main difference is that the output alphabet $\Gamma$ of the Mealy machine $\mathcal{M}$ is not $\mathcal{P}(\AP)$ but $\mathbb{R}^Y$.

\subsection{Active automata learning}\label{subsection:active_automata_learning}

\emph{Active automata learning} is a class of algorithms to construct an automaton
by a series of interactions between the \emph{learner} and a \emph{teacher}.
In L*~\cite{DBLP:journals/iandc/Angluin87} and TTT~\cite{DBLP:conf/rv/IsbernerHS14} algorithms, 
the learner constructs the minimum DFA $\mathcal{A}_{U}$ over $\Sigma$ recognizing the target language $U \subseteq \Sigma^*$
utilizing \emph{membership} and \emph{equivalence} questions to the teacher.

In a membership question, the learner asks if a word $w \in \Sigma^*$ is a member of $U$, \ie{} $w \in U$.
In an equivalence question, the learner asks if a candidate DFA $\mathcal{A}$ recognizes the target language $U$, \ie{} $\mathcal{L(A)} = U$.
In the equivalence question, if we have $\mathcal{L(A)} \neq U$, the teacher returns a word $w'$ satisfying $w' \in \mathcal{L(A)} \mathrel{\triangle} U$ as a witness of $\mathcal{M} \neq \Mealy$, where $\mathcal{L(A)} \mathrel{\triangle} U$ is the symmetric difference, \ie{} $\mathcal{L(A)} \mathrel{\triangle} U = (\mathcal{L(A)} \setminus U) \cup (U \setminus \mathcal{L(A)})$.
We note that a Mealy machine $\mathcal{M}$ can also be learned similarly. See \eg~\cite{DBLP:conf/sfm/SteffenHM11}.

\begin{algorithm}[tbp]
  \caption[]{L*-style active automata learning}%
  \label{algorithm:L*styleLearning}
  \DontPrintSemicolon{}
  \newcommand{\myCommentFont}[1]{\texttt{\footnotesize{#1}}}
  \SetCommentSty{myCommentFont}
  \SetKwFunction{FAskMembershipQ}{askMembershipQuestion}
  \SetKwFunction{FConstructCandidate}{constructCandidateAutomaton}
  \Input{A teacher $T$ that answers membership and equivalence questions of target language $U$}
  \Output{The minimum DFA $\mathcal{A}$ satisfying $U = \mathcal{L(A)}$}

  $\mathrm{observations} \gets \emptyset$\;
  \While {$\top$} {
    \tcp{Candidate generation phase}
    \While {$\exists w.$ we need to know if $w \in U$ to construct a candidate automaton~$\mathcal{A}$ from $\mathrm{observations}$} {\label{algorithm:L*style:AutomatonConstruction}
      add $(w, \FAskMembershipQ(T, w))$ to $\mathrm{observations}$\;
    }
    $\mathcal{A} \gets \FConstructCandidate(\mathrm{observations})$\label{algorithm:L*style:AutomatonConstructionEnd}\;\;
    \tcp{Equivalence testing phase}
    \If {$U = \mathcal{L}(\mathcal{A})$ by equivalence question} {\label{algorithm:L*style:EquivalenceTesting}
      \KwReturn $\mathcal{A}$\;
    } \Else {
      $w \gets$ a witness of $U \neq \mathcal{L}(\mathcal{A})$\label{algorithm:L*style:WitnessReturn}\;
      add $(w, \FAskMembershipQ(T, w))$ to $\mathrm{observations}$\;
    }\label{algorithm:L*style:EquivalenceTestingEnd}
  }
\end{algorithm}

\cref{algorithm:L*styleLearning} outlines the L*-style active automata learning algotithm.
In L*-style active automata learning, the learning process proceeds in two repetitive phases: candidate generation and equivalence testing. First, in the candidate generation phase (\crefrange{algorithm:L*style:AutomatonConstruction}{algorithm:L*style:AutomatonConstructionEnd}), the learner asks several membership questions to the teacher and constructs a candidate automaton. Once the automaton is constructed, the learning process proceeds to the equivalence testing phase (\crefrange{algorithm:L*style:EquivalenceTesting}{algorithm:L*style:EquivalenceTestingEnd}). The learner asks an equivalence question, and if the teacher returns a witness of inequivalence in \cref{algorithm:L*style:WitnessReturn}, the learning process returns to the first phase.

For any (even \emph{black-box}) system $\mathcal{M}$, we can learn a Mealy machine $\Mealy$ approximating the system behavior by implementing a teacher answering membership and equivalence questions.
It is usually easy to answer a membership question---we can answer it by executing $\mathcal{M}$.
In contrast, it is not straightforward to answer an equivalence question if the internal structure of the system $\mathcal{M}$ is unknown.
When we know the size of the automaton to represent the system $\mathcal{M}$, we can utilize 
conformance testing with the correctness guarantee, such as W-method~\cite{DBLP:journals/tse/Chow78} and Wp-method~\cite{DBLP:journals/tse/FujiwaraBKAG91}.
However, we usually do not know the size of such an automaton, and thus, we need an approximate method to test the equivalence of the system $\mathcal{M}$ under learning and the candidate automaton $\Mealy$, \eg{} by random testing and mutation testing~\cite{DBLP:journals/jar/AichernigT19}.
We note that, in general, these equivalence testing methods execute the system $\mathcal{M}$ for many times, and tend to be time-consuming when the system execution is expensive.

\subsection{Black-box checking}
\emph{Black-box checking (BBC)}~\cite{DBLP:conf/forte/PeledVY99} is a testing method that combines active automata learning and model checking to test if the given black-box system $\mathcal{M}$ satisfies its specification $\varphi$.
Given a black-box system $\mathcal{M}$ over an input alphabet $\Sigma$ and an output alphabet $\mathcal{P}(\AP)$, and a safety LTL formula $\varphi$,
BBC deems $\mathcal{M} \models \varphi$ or returns a counterexample $\sigma \in (\Sigma \times \mathcal{P}(\AP))^*$ such that 
for any $\sigma' \in {(\Sigma \times \mathcal{P}(\AP))}^{\omega}$ satisfying $\sigma \cdot \sigma' \in \mathcal{L(M)}$,
we have $\mathbf{pr_2}(\sigma \cdot \sigma') \not\models \varphi$.

\cref{fig:bbc-flow} outlines the workflow of BBC.\@
BBC combines L*-style active automata learning in \cref{algorithm:L*styleLearning} and model checking.
More precisely,
candidate generation phase (\crefrange{algorithm:L*style:AutomatonConstruction}{algorithm:L*style:AutomatonConstructionEnd} in \cref{algorithm:L*styleLearning}) corresponds to (A) in \cref{fig:bbc-flow},
equivalence testing phase of active automata learning (\crefrange{algorithm:L*style:EquivalenceTesting}{algorithm:L*style:EquivalenceTestingEnd} in \cref{algorithm:L*styleLearning}) corresponds to (D) in \cref{fig:bbc-flow}, and
model checking is used in (B) in \cref{fig:bbc-flow}.

First, we learn a Mealy machine $\Mealy$ approximating the behavior of the system $\mathcal{M}$ under test ((A) in \cref{fig:bbc-flow}).
We learn such a Mealy machine $\Mealy$ by the candidate generation of active automata learning (\crefrange{algorithm:L*style:AutomatonConstruction}{algorithm:L*style:AutomatonConstructionEnd} in \cref{algorithm:L*styleLearning}).
We note that the behavior of the learned Mealy machine $\Mealy$ may be different from that of the system $\mathcal{M}$ under test.

Then, we check if we have $\Mealy \models \varphi$ by model checking ((B) in \cref{fig:bbc-flow}).
If $\Mealy \not\models \varphi$ holds, the model checker returns a witness $\sigma \in {(\Sigma \times \mathcal{P}(\AP))}^*$ of $\Mealy \not\models \varphi$, and
we feed $\sigma$ to the system $\mathcal{M}$ under test to check if $\sigma$ is a witness of $\mathcal{M} \not\models \varphi$ ((C) in \cref{fig:bbc-flow}).
If $\sigma$ witnesses $\mathcal{M} \not\models \varphi$, we conclude that $\mathcal{M} \not\models \varphi$ holds, and BBC returns $\sigma$ as a counterexample.
Otherwise, since we have $\sigma \in \mathcal{L}^{\mathit{fin}}(\Mealy)$ and $\sigma \not\in \mathcal{L}^{\mathit{fin}}(\mathcal{M})$, $\sigma$ differentiates $\Mealy$ and $\mathcal{M}$, and we use $\sigma$ to refine the learned Mealy machine $\Mealy$.

If $\Mealy \models \varphi$ holds in the model-checking step ((B) in \cref{fig:bbc-flow}), we test if the behavior of $\Mealy$ and $\mathcal{M}$ are similar enough by equivalence testing of active automata learning ((D) in \cref{fig:bbc-flow}).
If we find an input $\sigma$ that differentiates $\mathcal{M}$ and $\Mealy$, we use $\sigma$ to refine the learned Mealy machine $\Mealy$.
Otherwise, we deem that $\Mealy$ and $\mathcal{M}$ are equivalent, and BBC returns $\mathcal{M} \models \varphi$.

\paragraph{BBC for CPSs}
To apply BBC to test a CPS $\mathcal{M}$, we need a finite abstraction of the real-valued input and output of $\mathcal{M}$.
Following~\cite{DBLP:conf/hybrid/Waga20}, we utilize input and output mappers $\mathcal{I}$ and $\mathcal{O}$ to bridge the real values for the CPS execution and the finite values for the BBC.
For a CPS model $\mathcal{M}$ over $X$ and $Y$, we fix the abstract input alphabet $\Sigma$ and the atomic propositions $\AP$, 
and define an input mapper $\mathcal{I} : \Sigma \to \mathbb{R}^X$ assigning one valuation of the input signal to each $a \in \Sigma$ and an output mapper $\mathcal{O} : \mathbb{R}^Y \to \mathcal{P}(\AP)$ assigning a set of atomic propositions to each valuation of the output signal.
Typically, $\Sigma$ is a finite subset of $\mathbb{R}^{X}$ and $\mathcal{I}$ is the canonical injection, and $\AP$ is a set of predicates over $Y$ and $\mathcal{O}$ assigns their satisfaction.

\section{BBC enhanced via model checking with strengthened LTL formulas}\label{section:our_method}

In this section, we show how we optimize BBC utilizing model checking with strengthened LTL formulas.
\cref{fig:our-workflow} shows the workflow of our enhanced BBC.
The high-level strategy is to reduce the number of the equivalence testing ((D) in \cref{fig:our-workflow}) via model checking with a strengthened LTL formula $\psi$ (($\mathrm{B}'$) and ($\mathrm{C}'$) in \cref{fig:our-workflow}).
Since, one equivalence test consists of many executions of the system $\mathcal{M}$ under test,
equivalence testing tends to be time-consuming if each execution of $\mathcal{M}$ is expensive.
In contrast, in BBC, the size of the learned Mealy machine $\Mealy$ tends to be small, and the model checking may be relatively fast.
Overall, the workflow in \cref{fig:our-workflow} may be more efficient than the original workflow of BBC in \cref{fig:bbc-flow}, which we experimentally confirm in \cref{section:experiment}.

\subsection{Strengthening relation of LTL formulas}

To formalize our strengthening of LTL formulas,
we define the strengthening relation ${\rightarrowtail} \subseteq \LTL \times \LTL$ over LTL formulas.
Given an LTL formula $\varphi$, we strengthen it to another LTL formula $\psi$ satisfying $\varphi \rightarrowtail \psi$.
The syntactic definition of ${\rightarrowtail}$ is suitable for the generation of the strengthened LTL formulas.

\begin{definition}
 [Strengthening relation of LTL formulas]\label{def:stl-str}
 For LTL formulas $\varphi, \psi$, ${\rightarrowtail} \subseteq \LTL \times \LTL$ is the minimum relation satisfying the following.

 \begin{enumerate}
  \item\label{rule:or} For any $\mu, \nu \in \LTL$, we have $(\mu \lor \nu) \rightarrowtail (\mu \land \nu)$.
  \item For any $\mu \in \LTL$, we have $\Evt \mu \rightarrowtail \Glb \Evt \mu$.
  \item For any $\mu \in \LTL$, we have $\Glb \Evt \mu \rightarrowtail \Evt \Glb \mu$.
  \item For any $\mu \in \LTL$, we have $\Evt \Glb \mu \rightarrowtail \Glb \mu$.
  \item For any $\mu \in \LTL$ and for any indices $i, j \in \mathbb{N} \cup \{\infty\}$ satisfying $i < j$, we have $\Evt_{[i, j)} \mu \rightarrowtail \Glb_{[i, j)} \mu$.
  \item For any $\mu, \nu \in \LTL$, we have $(\mu \Until \nu) \rightarrowtail (\Glb \mu \land \Glb \Evt \nu)$.
  \item\label{rule:evt-interval}
  For any $\mu \in \LTL$ and for any indices $i, j, i', j' \in \mathbb{N} \cup \{\infty\}$ satisfying $[i, j) \supsetneq [i', j')$, we have $\Evt_{[i,j)} \mu \rightarrowtail \Evt_{[i',j')} \mu$.
  \item\label{rule:neg} For any $\mu, \nu \in \LTL$, if we have $\nu \rightarrowtail \mu$, we have $\neg \mu \rightarrowtail \neg \nu$.
  \item For any $\mu, \mu', \nu \in \LTL$ satisfying $\mu \rightarrowtail \mu'$, we have $(\mu \lor \nu) \rightarrowtail (\mu' \lor \nu)$.
  \item\label{rule:or-subexp-right} For any $\mu, \nu, \nu' \in \LTL$ satisfying $\nu \rightarrowtail \nu'$, we have $(\mu \lor \nu) \rightarrowtail (\mu \lor \nu')$.
  \item For any $\mu, \nu \in \LTL$ satisfying $\mu \rightarrowtail \nu$,  we have $\Next \mu \rightarrowtail \Next \nu$.
  \item For any $\mu, \nu, \nu' \in \LTL$ satisfying $\nu \rightarrowtail \nu'$ and for any indices $i, j \in \mathbb{N} \cup \{\infty\}$ satisfying $i < j$, we have $(\mu \Until[[i, j)] \nu) \rightarrowtail (\mu \Until[[i, j)] \nu')$.
  \item For any $\varphi, \mu, \psi \in \LTL$ satisfying $\varphi \rightarrowtail \mu$ and $\mu \rightarrowtail \psi$, we have $\varphi \rightarrowtail \psi$.
 \end{enumerate}
\end{definition}

We note that for the other operators than the ones in \cref{definition:LTL}, ${\rightarrowtail}$ is defined using their definition as the syntactic abbreviation.

\begin{example}
 For any $p \in \mathbf{AP}$, we have $\Glb_{[0, 2)} p \rightarrowtail \Glb_{[0, 10)} p$.
 This is because, by condition~\ref{rule:evt-interval} of \cref{def:stl-str}, we have $\Evt_{[0, 10)} \neg p \rightarrowtail \Evt_{[0, 2)} \neg p$.
 By applying condition~\ref{rule:neg} of \cref{def:stl-str}, we obtain $\neg \Evt_{[0, 2)} \neg p \rightarrowtail \neg \Evt_{[0, 10)} \neg p$.
 By definition of the syntactic abbreviation, $\neg \Evt_{[0, 2)} \neg p \rightarrowtail \neg \Evt_{[0, 10)} \neg p$ is equivalent to $\Glb_{[0, 2)} p \rightarrowtail \Glb_{[0, 10)} p$. 
\end{example}

We have the following correctness by induction. \LongVersion{The proof is in \cref{apdx:proof:rewrite-correct}.}

\begin{theorem}
  [Correctness of the strengthening relation]\label{th:rewrite-correct}
  For any LTL formulas $\varphi$ and $\psi$ satisfying $\varphi \rightarrowtail \psi$, $\psi$ is stronger than $\varphi$, \ie{} for any $\pi \in (\mathcal{P}(\AP))^\omega$ and $k \in \mathbb{N}$, $(\pi, k) \models \varphi$ implies $(\pi, k) \models \psi$.
 \qed{}
\end{theorem}

\begin{example}
 Let $\varphi_{\mathit{example}} = p_1 \lor \Evt_{[0, 10)} p_2$, with $p_1, p_2 \in \AP$.
 By condition~\ref{rule:or} of \cref{def:stl-str}, we have $(p_1 \lor \Evt_{[0, 10)} p_2) \rightarrowtail (p_1 \land \Evt_{[0, 10)} p_2)$.
 Therefore, $p_1 \land \Evt_{[0, 2)} p_2$ is one of the candidates in the strengthening of $\varphi_{example}$.
 By conditions~\ref{rule:evt-interval} and~\ref{rule:or-subexp-right} of \cref{def:stl-str}, 
 we have
 $\Evt_{[0, 10)} p_2 \rightarrowtail \Evt_{[0, 5)} p_2$, and 
 $(p_1 \lor \Evt_{[0, 10)} p_2) \rightarrowtail (p_1 \lor \Evt_{[0, 5)} p_2)$.
 Therefore, $p_1 \lor \Evt_{[0, 5)} p_2$ is another candidate in the strengthening of $\varphi_{example}$.
 We note that by condition~\ref{rule:evt-interval} of \cref{def:stl-str}, 
 we have $\Evt_{[0, 10)} p_2 \rightarrowtail \Evt_{[i', j')} p_2$ for any $[i',j') \subsetneq [0,10)$, and
 in the strengthening,
 we have many candidates that are different only in the interval in their temporal operator.
 For example,
 $p_1 \lor \Evt_{[0, 8)} p_2$, $p_1 \lor \Evt_{[0, 3)} p_2$, and $p_1 \lor \Evt_{[0, 1)} p_2$ are the candidates in the strengthening of $\varphi_{\mathit{example}}$.
\end{example}

\subsection{BBC enhanced via model checking with strengthened formulas}

We present how we enhance BBC utilizing model checking with strengthened LTL formulas. 
In this section, we show the high-level scheme of our enhancement and, in \cref{subsection:detail_of_our_enhancement}, we explain the design choice in our implementation.
We fix the system $\mathcal{M}$ under test and the specification $\varphi \in \LTL$.

\cref{fig:our-workflow} outlines our enhanced BBC scheme.
When we have $\Mealy \models \varphi$ in (B) of \cref{fig:our-workflow},
before conducting the equivalence testing ((D) of \cref{fig:our-workflow}),
we try to find a witness of $\mathcal{M} \neq \Mealy$ by a model checking with an LTL formula $\psi$ satisfying $\varphi \rightarrowtail \psi$ (($\mathrm{B}'$) of \cref{fig:our-workflow}).
Since $\Mealy \nvDash \varphi$ implies $\Mealy \nvDash \psi$,
by model checking, we have more chance to obtain a witness $\sigma$ of $\Mealy \not\models \psi$ than that of $\Mealy \not\models \varphi$.
When $\psi$ is much stronger than $\varphi$,
the witness $\sigma$ of $\Mealy \not\models \psi$ is also a witness of $\mathcal{M} \not\models \psi$.
In such a case, $\sigma$ does not differentiate $\Mealy$ and $\mathcal{M}$, and thus, we cannot use $\sigma$ to refine $\Mealy$.
Nevertheless, we claim that if the LTL formula $\varphi$ is strengthened appropriately, we can often refine $\Mealy$ by such a witness $\sigma$.
Moreover, the refinement by such a witness $\sigma$ tends to lead to a Mealy machine useful for falsification of $\varphi$, which is observed in our experiment result in \cref{section:experiment}.

\begin{algorithm}[tb]
  \caption{BBC enhanced via model checking with strengthened LTL formulas}%
  \label{algorithm:BBC}
  \DontPrintSemicolon{}
  \newcommand{\myCommentFont}[1]{\texttt{\footnotesize{#1}}}
  \SetCommentSty{myCommentFont}
  \SetKwFunction{FStrSTL}{SyntacticStrengthening}
  \SetKwFunction{FGenCandidate}{GenCandidate}
  \SetKwFunction{FChooseSpec}{ChooseFml}
  \SetKwFunction{FConstructInitMealy}{ConstructInitialMealy}
  \SetKwFunction{FRefineMealy}{RefineMealy}
  \SetKwFunction{FisTimeout}{isTimeout}

  \Input{System $\mathcal{M}$ under test and an LTL formula $\varphi$}
  \Output{Returns $\top$ if BBC deems $\mathcal{M} \models \varphi$, otherwise, a witness $\sigma$ of $\mathcal{M} \not\models \varphi$ }
 $\Psi \gets \FGenCandidate(\varphi)$
 \tcp{Generate a subset $\Psi$ of $\{\psi \in \LTL \mid \varphi \rightarrowtail \psi \}$}
 \label{algorithm:BBC:StrSTL}
  $\Mealy \gets \FConstructInitMealy(\mathcal{M})$\;\label{algorithm:BBC:ConstructInitMealy}
  \Repeat{$\FisTimeout{}$} {\label{algorithm:BBC:main_loop:begin}
    \If {$\Mealy \nvDash \varphi$} {\label{algorithm:BBC:MCOriginalSpec}
      $\sigma \gets$ a witness of $\Mealy \nvDash \varphi$\;
      \If {$\sigma$ witnesses $\mathcal{M} \nvDash \varphi$} {\label{algorithm:BBC:Check_sigma_original}
        \KwReturn{$\sigma$}
      }
    } \Else {
      $\mathrm{foundWitness} \gets \bot$\;
      $\Psi_{\mathit{chosen}} \gets \FChooseSpec(\Psi)$\;\label{algorithm:BBC:AdaptiveSpec}
      \ForAll(\tcp*[h]{Try the strengthened specifications}){$\psi_i \in \Psi_{\mathit{chosen}}$} {\label{algorithm:BBC:Phi_forall}
        \If{$\Mealy \nvDash \psi_i$} {\label{algorithm:BBC:MCStrSpec}
          $\sigma \gets$ a witness of $\Mealy \not\models \psi_i$\;
          \If {$\sigma$ witnesses $\mathcal{M} \not\models \psi_i$} {\label{algorithm:BBC:Check_sigma_strengthened}
            remove $\psi_i$ from $\Psi$\;\label{algorithm:BBC:remove_varphi_i}
          } \Else(\tcp*[h]{$\sigma$ is a witness of $\Mealy \neq \mathcal{M}$}) {\label{algorithm:BBC:foundWitness}
            $\mathrm{foundWitness} \gets \top$\;
            \KwBreak\;
          }
        }
      }
      \If {$\mathrm{foundWitness} = \bot$} {
        \If {$\Mealy \simeq \mathcal{M}$ by equivalence testing} {\label{algorithm:BBC:equivalence_testing}
          \KwReturn{$\top$}
        } \Else {
          $\sigma \gets$ a witness of $\Mealy \neq \mathcal{M}$
        }
      }
    }
    $\Mealy \gets \FRefineMealy(\mathcal{M}, \sigma)$\;\label{algorithm:BBC:RefineMealy}
  }\label{algorithm:BBC:main_loop:end}
  \KwReturn{$\top$}
\end{algorithm}

\cref{algorithm:BBC} outlines our BBC enhanced via model checking with strengthened LTL formulas.
In \cref{algorithm:BBC:StrSTL}, 
we generate the candidates $\candidates$ of the strengthened LTL formulas used in the model checking.
After constructing the initial Mealy machine $\Mealy$ in \cref{algorithm:BBC:ConstructInitMealy},
we conduct model checking of $\Mealy$ with $\varphi$.
When we have $\Mealy \nvDash \varphi$ (\cref{algorithm:BBC:MCOriginalSpec}),
we obtain a witness $\sigma$ of $\Mealy \nvDash \varphi$ and check if $\sigma$ also witnesses $\mathcal{M} \nvDash \varphi$ by running $\mathcal{M}$ with $\sigma$ as the input (\cref{algorithm:BBC:Check_sigma_original}).
When $\sigma$ also witnesses $\mathcal{M} \nvDash \varphi$, we return $\sigma$ as a result of BBC.
Otherwise, we use $\sigma$ to refine the leaned Mealy machine $\Mealy$ (\cref{algorithm:BBC:RefineMealy}).

When we have $\Mealy \models \varphi$, we look for an input $\sigma$ to refine $\Mealy$.
In the original BBC in \cref{fig:bbc-flow}, we try the equivalence testing to find such $\sigma$,
In contrast, in order to reduce the number of the equivalence testing, we conduct model checking of $\Mealy$ with some of the LTL formulas $\psi \in \Psi$
before trying the equivalence testing.
The strengthened LTL formulas $\Psi_{\mathit{chosen}}$ is chosen by a function $\FChooseSpec$.
Although the stronger LTL formulas should be chosen before the weaker ones, $\FChooseSpec$ can be an arbitrary function to choose a finite set of the strengthened specifications $\Psi_{\mathit{chosen}}$ from $\Psi$.
We note that the choice of $\FGenCandidate$ and $\FChooseSpec$ defines the granularity of the strengthening of $\varphi$ used in the model checking, which may affect the effectiveness of our enhancement.

For each LTL formula $\psi_i \in \Psi_{\mathit{chosen}}$,
we check if $\Mealy \nvDash \psi_i$ holds by model checking in \cref{algorithm:BBC:Phi_forall}.
When $\Mealy \nvDash \psi_i$ holds (\cref{algorithm:BBC:MCStrSpec}), we obtain a witness $\sigma$ of $\Mealy \nvDash \psi_i$.
Then, we check if $\sigma$ also witnesses $\mathcal{M} \nvDash \psi_i$ by running $\mathcal{M}$ with $\sigma$ as input (\cref{algorithm:BBC:Check_sigma_strengthened}).
When $\sigma$ also witnesses $\mathcal{M} \nvDash \varphi_i$,
we remove $\psi_i$ from $\Psi$ in \cref{algorithm:BBC:remove_varphi_i}.
Otherwise, we use $\sigma$ to refine the learned Mealy machine $\Mealy$ in \cref{algorithm:BBC:RefineMealy}.

When for any $\psi_i \in \Psi_{\mathit{chosen}}$, we can not find $\sigma$ to refine $\Mealy$,
we fallback to the normal loop of the BBC.
Namely,
we use equivalence testing to find a witness $\sigma$ of $\mathcal{M} \neq \Mealy$ in \cref{algorithm:BBC:equivalence_testing}.
When equivalence testing deems $\Mealy$ and $\mathcal{M}$ are equivalent, we return $\top$ as the result of BBC.
Otherwise, equivalence testing returns a witness $\sigma$ of $\Mealy \neq \mathcal{M}$, and we use $\sigma$ to refine $\Mealy$ (\cref{algorithm:BBC:RefineMealy}).

\subsection[]{\FGenCandidate and \FChooseSpec in our implementation}\label{subsection:detail_of_our_enhancement}

\begin{algorithm}[t]
  \caption[]{The candidate generation \FGenCandidate in our implementation, where $N \in \mathbb{N}$ is the bound of the time horizon}%
  \label{algorithm:CandidateGeneration}
  \DontPrintSemicolon{}
  \newcommand{\myCommentFont}[1]{\texttt{\footnotesize{#1}}}
  \SetCommentSty{myCommentFont}
  \SetKwFunction{FCandOpSpecs}{GenNoIntFml}
  \SetKwFunction{FCandIntvSpecs}{GenIntFml}
  \Fn{\FGenCandidate{$\varphi$}} {
     \Input{An LTL formula $\varphi$}
     \Output{The strengthened LTL formulas $\Psi$ used in \cref{algorithm:BBC}}
     $\noIntCandidates \gets \text{\FCandOpSpecs{$\varphi$}}$ \tcp{Strengthen the operators without intervals}
     $\intCandidates \gets \text{\FCandIntvSpecs{$\varphi$}}$  \tcp{Strengthen the operators with intervals}
     \KwReturn $\noIntCandidates \cup \intCandidates$\;
  }
  \Fn{\FCandIntvSpecs{$\varphi$}} {
    $\intCandidates \gets \emptyset$\;
    \Switch{the syntactic structure of $\varphi$} {
      \Case{$\varphi = \Glb_{[i,j)} \mu$} {\label{algorithm:CandIntvSpecs:caseGlb}
        $i' \gets 0$;\,\, $j' \gets \infty$\;\label{algorithm:CandIntvSpecs:GlbInitIntv}
        \While{$[i,j) \subsetneq [i',j')$}{
          $\intCandidates \gets \intCandidates \cup \{ \Glb_{[i',j')} \mu \}$\;
          \lIf{$i > i'$} {
            $i' \gets \lceil \frac{i + i'}{2} \rceil$;\,\, $j' \gets N$\label{algorithm:CandIntvSpecs:GlbUpdLower}
          }
            \lElse{
              $j' \gets \lfloor \frac{j + j'}{2} \rfloor$\label{algorithm:CandIntvSpecs:GlbUpdUpper}
          }
        }
      }
      \Case{$\varphi = \Evt_{[i,j)} \mu$} {
        $\intCandidates \gets \FCandIntvSpecs(\Glb_{[i,i+1)} \mu)$\;
        $i' \gets i$;\,\, $j' \gets i + 1$\;
        \While{$[i,j) \supsetneq [i',j')$}{
          $\intCandidates \gets \intCandidates \cup \{\Evt_{[i',j')} \mu\}$\;
          \lIf{$i < i'$} {
            $i' \gets \lfloor \frac{i + i'}{2} \rfloor$
          }
          \lElse{
            $j' \gets \lceil \frac{j + j'}{2} \rceil$
          }
        }
      }
      \Case{$\varphi = \Glb \mu$} {
        $\intCandidates \gets \{ \Glb \mu' \mid \mu' \in \text{\FCandIntvSpecs{$\mu$}} \}$\;
      }
      \Case{$\varphi = \mu \lor \nu$} {
        $\intCandidates \gets \{ \mu' \lor \nu \mid \mu' \in \text{\FCandIntvSpecs{$\mu$}} \} \cup \{ \mu \lor \nu' \mid \nu' \in \text{\FCandIntvSpecs{$\nu$}} \}$\;
      }
      \Case{$\varphi = \mu \land \nu$} {
        $\intCandidates \gets \{ \mu' \land \nu \mid \mu' \in \text{\FCandIntvSpecs{$\mu$}} \} \cup \{ \mu \land \nu' \mid \nu' \in \text{\FCandIntvSpecs{$\nu$}} \}$\;
      }
    }
    \KwReturn $\intCandidates$\;
  }
\end{algorithm}

\begin{algorithm}[tbp]
  \caption{Candidate generation by strengthening the operators without intervals}
  \label{algorithm:CandidateOperatorSpecs}
  \DontPrintSemicolon{}
  \newcommand{\myCommentFont}[1]{\texttt{\footnotesize{#1}}}
  \SetCommentSty{myCommentFont}
  \Input{An LTL formula $\varphi$}
  \Output{A queue $\noIntCandidates$ of LTL formulas that are obtained by strengthening the operators without intervals in $\varphi$}

  \Fn{\FCandOpSpecs{$\varphi$}}{
    $\noIntCandidates \gets ()$ \tcp*[h]{$\noIntCandidates$ is a queue of strengthened specs}\;
    \Switch{the form of $\varphi$} {
      \Case{$\varphi = \mu \lor \nu$}{
        \KwPush{} $\mu \land \nu$ \KwTo{} $\noIntCandidates$\;
        \ForAll{$\mu' \in \FCandOpSpecs(\mu)$}{
          \KwPush{} $\mu' \lor \nu$ \KwTo{} $\noIntCandidates$\;
        }
        \ForAll{$\nu' \in \FCandOpSpecs(\nu)$}{
          \KwPush{} $\mu \lor \nu'$ \KwTo{} $\noIntCandidates$\;
        }
      }
      \Case{$\varphi = \Evt \mu$} {
        \KwReturn ($\Glb \mu$, $\Evt\Glb \mu$, $\Glb\Evt \mu$, $\Evt \mu$)\;
      }
      \Case{$\varphi = \mu \Until \nu$} {
        \KwReturn ($\Glb \mu \land \Glb \nu$, $\Glb \mu \land \Evt\Glb \nu$, $\Glb \mu \land \Glb\Evt \nu$)\;
      }
      \Case{$\varphi = \mu \land \nu$} {
        \ForAll{$\mu' \in \FCandOpSpecs(\mu)$} {
          \KwPush{} $\mu' \land \nu$ \KwTo{} $\noIntCandidates$\;
        }
        \ForAll{$\nu' \in \FCandOpSpecs(\nu)$}{
          \KwPush{} $\mu \land \nu'$ \KwTo{} $\noIntCandidates$\;
        }
      }
      \Case{$\varphi = \Glb \mu$} {
        \ForAll{$\mu' \in \FCandOpSpecs(\mu)$}{
          \KwPush{} $\Glb \mu'$ \KwTo{} $\noIntCandidates$\;
        }
      }
    }
    \KwReturn $\noIntCandidates$\;
  }
\end{algorithm}

\cref{algorithm:CandidateGeneration} shows our candidate generation algorithm \FGenCandidate.
The candidates $\candidates$ of the strengthened LTL formulas consists of $\intCandidates$ and $\noIntCandidates$\footnote{More precisely, $\noIntCandidates$ is a queue and its FIFO order is used in $\FChooseSpec$ in \cref{algorithm:ChooseSpec}. 
}:
$\intCandidates$ and $\noIntCandidates$ are obtained by strengthening the operators with and without intervals.
They are constructed by \FCandIntvSpecs and \FCandOpSpecs (in \cref{algorithm:CandidateOperatorSpecs}), respectively.
Moreover, we remove $\psi_i$ from $\intCandidates$ or $\noIntCandidates$ when $\psi_i$ is removed from $\candidates$ in \cref{algorithm:BBC:remove_varphi_i} of \cref{algorithm:BBC}.

First, we use \FCandOpSpecs to construct $\noIntCandidates \subseteq \{ \psi \in \LTL \mid \varphi \rightarrowtail \psi \}$ that is constructed by inductively strengthening the operators without intervals.
For example, for $\varphi = (\Glb_{[2,6)} p) \lor \Evt q$, we have $\text{\FCandOpSpecs{$\varphi$}} = \{ (\Glb_{[2,6)} p) \land \Evt q, (\Glb_{[2,6)} p) \lor \Glb q, (\Glb_{[2,6)} p) \lor \Evt \Glb q, (\Glb_{[2,6)} p) \lor \Glb \Evt q\}$.
We note that for any LTL formula $\varphi$, \FCandOpSpecs{$\varphi$} is a finite set.

Then, we use $\FCandIntvSpecs$ to construct a finite set $\intCandidates$ of LTL formulas by modifying the ``Eventually'' and ``Globally'' operators with intervals in $\varphi$.
We employ heuristics to take the midpoint of the lower or upper bound when shrinking the interval.
For example, let $\varphi = (\Glb_{[2,6)} p) \lor \Evt q$ and the bound $N$ of the time horizon be $N = 30$. 
We start from $[i',j') = [0, \infty)$ (in \cref{algorithm:CandIntvSpecs:GlbInitIntv} of \cref{algorithm:CandidateGeneration}) and repeatedly update the lower bound $i'$ to the midpoint of $i$ and $i'$ (\cref{algorithm:CandIntvSpecs:GlbUpdLower}) to generate an LTL formula with it.
Namely, we generate $(\Glb_{[0,\infty)} p) \lor \Evt q$, $(\Glb_{[1,30)} p) \lor \Evt q$, and $(\Glb_{[2,30)} p) \lor \Evt q$.
Once we have $i = i'$, we repeatedly update the upper bound $j'$ to the midpoint of $j$ and $j'$ (\cref{algorithm:CandIntvSpecs:GlbUpdUpper}), and use $[i',j')$ for the LTL generation.
Namely, we generate $(\Glb_{[2,18)} p) \lor \Evt q$, $(\Glb_{[2,12)} p) \lor \Evt q$, $(\Glb_{[2,9)} p) \lor \Evt q$, and $(\Glb_{[2,7)} p) \lor \Evt q$.
By this construction, we have finer-grained strengthening when the strengthened formula is closer to the original formula while ignoring many strengthened formulas far from the original one for efficiency.

\begin{algorithm}[tbp]
  \caption[]{Our implementation of \FChooseSpec}%
  \label{algorithm:ChooseSpec}
  \DontPrintSemicolon{}
  \newcommand{\myCommentFont}[1]{\texttt{\footnotesize{#1}}}
  \SetCommentSty{myCommentFont}
  \Input{A set $\candidates$ of the candidates of the strengthened LTL formulas consists of $\intCandidates$ and $\noIntCandidates$}
  \Output{A set $\Psi_{\mathit{chosen}}$ of LTL formulas chosen from $\candidates$}

  $\Psi_{\mathit{chosen}} \gets \emptyset$\;
  $\noIntCandidates['] \gets \noIntCandidates$\;
  \tcp{Find the first formula in $\noIntCandidates$ with no stronger formulas in $\noIntCandidates$}
  \While {$\noIntCandidates['] \neq \emptyset$} {
    \KwPop{} $\psi$ \KwFrom{} $\noIntCandidates[']$\;
    \If{$\forall \psi' \in \noIntCandidates['].\, \psi \not\succeq \psi'$} {
      $\Psi_{\mathit{chosen}} \gets \Psi_{\mathit{chosen}} \cup \{\psi\}$\;
      \KwBreak{}
    }
  }
  $\Psi_{\mathit{chosen}} \gets \Psi_{\mathit{chosen}} \cup \{ \psi \in \intCandidates \mid \forall \psi' \in \intCandidates.\, \psi \not\succeq \psi' \}$\;
  \KwReturn $\Psi_{\mathit{chosen}}$\;
\end{algorithm}

In $\FChooseSpec$ (in \cref{algorithm:ChooseSpec}), we take one of the strongest LTL formulas in $\noIntCandidates$ and take all the strongest LTL formulas in $\intCandidates$.
We note that the strength of LTL formulas is a strict partial order, and there may be multiple strongest specifications.

\section{Experiment}\label{section:experiment}

We conducted experiments to evaluate the efficiency of our BBC enhanced by model checking with strengthened LTL formulas.
We compared our method with a tool FalCAuN~\cite{DBLP:conf/hybrid/Waga20} for robustness-guided BBC for CPSs.
We implemented a prototype tool based on FalCAuN in Java
\footnote{Our implementation is publicly available in \url{https://github.com/MasWag/FalCAuN/releases/tag/RV2021}.}.

\subsection{Experiment setup}
As the CPS $\mathcal{M}$ under test, we used the Simulink model of an automatic transmission system~\cite{DBLP:conf/cpsweek/HoxhaAF14},
one of the standard models in the falsification literature.
Given a 2-dimensional signal of the throttle and the brake, the automatic transmission model $\mathcal{M}$ returns a 3-dimensional signal of the velocity $v$, the engine rotation $\omega$, and the gear $g$.
The range of the throttle and the brake are $[0, 100]$ and $[0, 325]$, respectively.
The domains of $v$ and $\omega$ are positive reals, and the domain of $g$ is $\{1,2,3,4\}$.
As the specification, we used the set of the STL formulas in \cref{table:specs}.
The STL formulas $\varphi_1$ and $\varphi_2$ are taken from~\cite{DBLP:journals/tcad/ZhangESAH18}, and $\varphi_3$-$\varphi_5$ are our original.
Since the length of the input and output signals in our experiment is less than 30,
we let the bound $N$ in \cref{algorithm:CandidateGeneration} be 30.


Since the input and the output of the system $\mathcal{M}$ under test are continuous, we cannot directly apply BBC for the falsification of $\mathcal{M}$.
In our experiments, we use the following discretization both in time and values.
For the discretization in time, we use fixed-interval sampling of every one second.
For the discretization of input values, we use the following 4 ($= 2 \times 2$) values: the throttle is either 0 or 100, and the brake is either 0 or 325.
For the discretization of output values, we use the coarsest atomic propositions $\AP$ that is a partition of the output range compatible with the inequalities in the STL formula in each benchmark.
For example, since the inequality constraints in the STL formula $\varphi_1$ are $v < 100$ and $v > 75$, the atomic propositions $\AP$ for $\varphi_1$ is $\{v \leq 75, 75 < v < 100, 100 \leq v\}$.

Among the optimization methods supported by FalCAuN to search for a counterexample in the equivalence testing, we use a genetic algorithm.
Due to the stochastic nature of a genetic algorithm, we executed each benchmark 50 times.
For each execution,
we measured the time and the number of the Simulink executions to falsify the STL formula.
We set the timeout of each execution to 4 hours.
We experimented on a Google Cloud Platform c2-standard-4 instance (4 vCPUs and 15.67GiB RAM). We used Debian 10 buster and MATLAB R2020b.

\begin{table}[t]
  \centering
  \caption{List of the STL formulas in our benchmarks}
  \label{table:specs}
  \setlength{\tabcolsep}{10pt}
  \begin{tabular}{c|c}
    & STL formula \\ \hline
    $\varphi_1$ & $\Glb_{[0, 26]} (v < 100) \lor \Glb_{[28, 28]} (v > 75)$ \\
    $\varphi_2$ & $\Glb ((\omega < 4770) \lor (\Glb_{[1, 1]} (\omega > 600)))$ \\
    $\varphi_3$ & $\Glb ((g > 3) \lor (\omega < 4775) \lor \Evt_{[0, 2]} (g > 3))$ \\
    $\varphi_4$ & $\Glb ((g > 2) \lor ((g < 2) \Until (v > 30)))$ \\
    $\varphi_5$ & $\Glb ((\Evt_{[0, 3]} (\omega < 4000)) \lor (\Evt_{[0, 3]} (v > 100)))$
  \end{tabular}
\end{table}

\subsection{Performance evaluation}

\begin{table}[t]
  \centering
  \caption{Summary of the experiment result of 50 executions for our benchmarks. The numbers $T/N$ in each cell at ``average'' and ``std. dev.'' columns are the time $T$ [min.] to falsify the specification and the number $N$ of Simulink executions to falsify the specification. The number $N$ in each cell at ``timeout'' column is the number $N$ of timeouts to falsify the specification. In this experiment, the timeout is 4 hours. For each benchmark $\varphi_i$, we highlight the best cell in average column in terms of the following order: $T/N$ is better than $T'/N'$ if and only if we have $T < T'$ or we have both $T = T'$ and $N < N'$. For each benchmark, the cells of the smallest number of timeouts is highlighted.}
  \label{table:result}
    \setlength{\tabcolsep}{4pt}
 \scriptsize
    \begin{tabular}{c||c|c|c||c|c|c}
      {} & \multicolumn{3}{c||}{Our method} & \multicolumn{3}{c}{Baseline (FalCAuN)} \\
      {} & average & std. dev. & timeout & average & std. dev. & timeout\\ \hline \hline
      $\varphi_1$ & \cellcolor{green!20} 19.29 / 6664.7 & 7.16 / 1962.7 & \cellcolor{green!20} 0 & 26.70 / 9471.0 & 15.19 / 5412.2 & \cellcolor{green!20} 0\\
      $\varphi_2$ & \cellcolor{green!20} 54.89 / 19066.1 & 42.38 / 13609.3 & \cellcolor{green!20} 5 & 78.71 / 27362.6 & 57.85 / 18761.1 & 13\\
      $\varphi_3$ & \cellcolor{green!20} 16.43 / 6068.8 & 18.65 / 6622.2 & \cellcolor{green!20} 1 & 17.35 / 6306.3 & 25.60 / 8195.7 & \cellcolor{green!20} 1\\
      $\varphi_4$ & \cellcolor{green!20} 2.53 / 957.0 & 1.08 / 478.6 & \cellcolor{green!20} 0 & 7.48 / 2323.5 & 5.40 / 1683.2 & \cellcolor{green!20} 0\\
      $\varphi_5$ & \cellcolor{green!20} 4.92 / 1785.4 & 2.07 / 803.5 & \cellcolor{green!20} 0 & 5.19 / 2003.4 & 2.31 / 904.5 & \cellcolor{green!20} 0
    \end{tabular}
\end{table}

\cref{table:result} shows the summary of the experiment results. Execution times are shown in minutes.
%
%
For each STL formula~$\varphi_i$, we observe that, on average, our method falsified~$\varphi_i$ in a shorter time than the baseline.
Moreover, on average, the number of Simulink executions of our method is smaller than that of baseline.
Furthermore, the number of timeouts of our method is smaller than or equal to that of the baseline.
Overall, the experiment results in \cref{table:result} suggest that model checking with strengthened STL formulas makes the BBC more efficient.

Although our method outperforms the baseline for all the STL formulas,
we also observe that the amount of acceleration differs among the formulas.
%
For $\varphi_4$, our method was about 66\% faster than the baseline, and acceleration was the largest. 
This is because our method generates four strengthened specifications by strengthening the ``Until'' operator in $\varphi_4$. They guided the learning of an automaton in BBC.\@
For $\varphi_1$ and $\varphi_2$, acceleration by our enhancement was about 27\% to 30\%, which is significant but not as much as the one for $\varphi_4$.
This is because our method generates many strengthened specifications by changing the interval of the ``Globally'' operators while model checking with them guided the Mealy machine learning in the BBC.\@
Although many specifications are generated by our specification strengthening,
the falsification of
 the original specifications in $\varphi_1$ and $\varphi_2$ is difficult and time consuming, 
 the overhead due to the model checking with many strengthened LTL formulas is not significant.

In contrast, for $\varphi_3$ and $\varphi_5$, our method was only about 5\% faster than the baseline.
For $\varphi_3$, by definition of the strengthening relation in \cref{def:stl-str}, falsification of most of the strengthened specifications requires the output signal to violate both $g > 3$ and $\omega < 4775$ (almost) at the same time, which is a falsification of a disjunctive specification and tends to be difficult~\cite{DBLP:journals/corr/abs-2012-00319}.
 Since falsification of most of the strengthened STL formulas is difficult, the improvement thanks to the model checking with them is limited.
One of the future directions to overcome this issue is enhancing genetic algorithm-based equivalence testing, \eg{} utilizing \emph{ranking}~\cite{DBLP:journals/corr/abs-2012-00319}.
Another direction is to strengthen the specification by modifying the thresholds to make the specification strengthening finer-grained.

For $\varphi_5$, since the original specification $\varphi_5$ is not difficult and we can falsify it relatively quickly,
we cannot ignore the overhead of model checking with the strengthened specifications.
For such a situation, possible future work is an improvement of the choice of the strengthened STL formulas, \eg{} by performing binary search on the strengthening of specifications to reduce the number of specifications to be model-checked.

\section{Conclusions and future work}\label{section:conclusions_and_future_work}


One of the issues in BBC for CPSs is its long execution time.
In particular, the execution time of the equivalence test tends to be the bottleneck because an equivalence test consists of many system executions and each execution of a CPS is time-consuming. 
To reduce the number of the equivalence tests, we proposed an enhancement of BBC via model checking with strengthened specifications.
By model checking with an LTL formula $\psi$ stronger than the original formula $\varphi$, we have more chance to obtain a witness of the violation, and 
such a witness tends to be helpful for the refinement of the learned Mealy machine $\Mealy$.
Our experiment result shows that our method accelerates BBC, and our method is up to 66 \% faster than the conventional BBC. 

When the complexity of the original LTL formula $\varphi$ is high, \eg{} containing many temporal operators, 
the number of the strengthened formulas tends to be huge.
In such a case,
 our current naive choice of the LTL formulas to be model checked, \ie{} $\FGenCandidate$ and $\FChooseSpec$ in \cref{algorithm:BBC}, 
may cause significant overhead.
One of the future works is to optimize such a choice of the model-checked formulas.
For example, 
utilizing a binary search on the strengthened formulas 
or 
rewriting multiple operators in the original formula at one time may reduce the number of the model checking execution. 
Another future work is to investigate other kinds of specification strengthening.
One example is to  change the threshold in the inequalities.
Optimization of the robustness-guided equivalence testing with recent falsification techniques, \eg~\cite{DBLP:journals/corr/abs-2012-00319}, is also future work.

\subsubsection*{Acknowledgments.}
This work is partially supported by JST ACT-X Grant No.\ JPMJAX200U, JSPS KAKENHI Grant Number 19H04084, and JST CREST Grant Number JPMJCR2012, Japan.

\bibliographystyle{splncs04}
\bibliography{main}

\begin{thebibliography}{10}
\providecommand{\url}[1]{\texttt{#1}}
\providecommand{\urlprefix}{URL }
\providecommand{\doi}[1]{https://doi.org/#1}

\bibitem{DBLP:journals/jar/AichernigT19}
Aichernig, B.K., Tappler, M.: Efficient active automata learning via mutation
  testing. J. Autom. Reason.  \textbf{63}(4),  1103--1134 (2019).
  \doi{10.1007/s10817-018-9486-0},
  \url{https://doi.org/10.1007/s10817-018-9486-0}

\bibitem{DBLP:journals/iandc/Angluin87}
Angluin, D.: Learning regular sets from queries and counterexamples. Inf.
  Comput.  \textbf{75}(2),  87--106 (1987). \doi{10.1016/0890-5401(87)90052-6},
  \url{https://doi.org/10.1016/0890-5401(87)90052-6}

\bibitem{DBLP:conf/tacas/AnnpureddyLFS11}
Annpureddy, Y., Liu, C., Fainekos, G.E., Sankaranarayanan, S.: S-taliro: {A}
  tool for temporal logic falsification for hybrid systems. In: Abdulla, P.A.,
  Leino, K.R.M. (eds.) Tools and Algorithms for the Construction and Analysis
  of Systems - 17th International Conference, {TACAS} 2011, Held as Part of the
  Joint European Conferences on Theory and Practice of Software, {ETAPS} 2011,
  Saarbr{\"{u}}cken, Germany, March 26-April 3, 2011. Proceedings. Lecture
  Notes in Computer Science, vol.~6605, pp. 254--257. Springer (2011).
  \doi{10.1007/978-3-642-19835-9\_21},
  \url{https://doi.org/10.1007/978-3-642-19835-9\_21}

\bibitem{DBLP:conf/cec/AugerH05}
Auger, A., Hansen, N.: A restart {CMA} evolution strategy with increasing
  population size. In: Proceedings of the {IEEE} Congress on Evolutionary
  Computation, {CEC} 2005, 2-4 September 2005, Edinburgh, {UK}. pp. 1769--1776.
  {IEEE} (2005). \doi{10.1109/CEC.2005.1554902},
  \url{https://doi.org/10.1109/CEC.2005.1554902}

\bibitem{DBLP:series/lncs/BartocciDDFMNS18}
Bartocci, E., Deshmukh, J.V., Donz{\'{e}}, A., Fainekos, G.E., Maler, O.,
  Nickovic, D., Sankaranarayanan, S.: Specification-based monitoring of
  cyber-physical systems: {A} survey on theory, tools and applications. In:
  Bartocci, E., Falcone, Y. (eds.) Lectures on Runtime Verification -
  Introductory and Advanced Topics, Lecture Notes in Computer Science, vol.
  10457, pp. 135--175. Springer (2018). \doi{10.1007/978-3-319-75632-5\_5},
  \url{https://doi.org/10.1007/978-3-319-75632-5\_5}

\bibitem{DBLP:conf/rv/CameronFMS15}
Cameron, F., Fainekos, G.E., Maahs, D.M., Sankaranarayanan, S.: Towards a
  verified artificial pancreas: Challenges and solutions for runtime
  verification. In: Bartocci, E., Majumdar, R. (eds.) Runtime Verification -
  6th International Conference, {RV} 2015 Vienna, Austria, September 22-25,
  2015. Proceedings. Lecture Notes in Computer Science, vol.~9333, pp. 3--17.
  Springer (2015). \doi{10.1007/978-3-319-23820-3\_1},
  \url{https://doi.org/10.1007/978-3-319-23820-3\_1}

\bibitem{DBLP:conf/dsd/CasagrandeP12}
Casagrande, A., Piazza, C.: Model checking on hybrid automata. In: 15th
  Euromicro Conference on Digital System Design, {DSD} 2012, Cesme, Izmir,
  Turkey, September 5-8, 2012. pp. 493--500. {IEEE} Computer Society (2012).
  \doi{10.1109/DSD.2012.87}, \url{https://doi.org/10.1109/DSD.2012.87}

\bibitem{DBLP:journals/tse/Chow78}
Chow, T.S.: Testing software design modeled by finite-state machines. {IEEE}
  Trans. Software Eng.  \textbf{4}(3),  178--187 (1978).
  \doi{10.1109/TSE.1978.231496}, \url{https://doi.org/10.1109/TSE.1978.231496}

\bibitem{DBLP:conf/cav/Donze10}
Donz{\'{e}}, A.: Breach, {A} toolbox for verification and parameter synthesis
  of hybrid systems. In: Touili, T., Cook, B., Jackson, P.B. (eds.) Computer
  Aided Verification, 22nd International Conference, {CAV} 2010, Edinburgh, UK,
  July 15-19, 2010. Proceedings. Lecture Notes in Computer Science, vol.~6174,
  pp. 167--170. Springer (2010). \doi{10.1007/978-3-642-14295-6\_17},
  \url{https://doi.org/10.1007/978-3-642-14295-6\_17}

\bibitem{DBLP:conf/formats/DonzeM10}
Donz{\'{e}}, A., Maler, O.: Robust satisfaction of temporal logic over
  real-valued signals. In: Chatterjee, K., Henzinger, T.A. (eds.) Formal
  Modeling and Analysis of Timed Systems - 8th International Conference,
  {FORMATS} 2010, Klosterneuburg, Austria, September 8-10, 2010. Proceedings.
  Lecture Notes in Computer Science, vol.~6246, pp. 92--106. Springer (2010).
  \doi{10.1007/978-3-642-15297-9\_9},
  \url{https://doi.org/10.1007/978-3-642-15297-9\_9}

\bibitem{ARCH20:ARCH_COMP_2020_Category_Report}
Ernst, G., Arcaini, P., Bennani, I., Donze, A., Fainekos, G., Frehse, G.,
  Mathesen, L., Menghi, C., Pedrielli, G., Pouzet, M., Yaghoubi, S., Yamagata,
  Y., Zhang, Z.: Arch-comp 2020 category report: Falsification. In: Frehse, G.,
  Althoff, M. (eds.) ARCH20. 7th International Workshop on Applied Verification
  of Continuous and Hybrid Systems (ARCH20). EPiC Series in Computing, vol.~74,
  pp. 140--152. EasyChair (2020). \doi{10.29007/trr1},
  \url{https://easychair.org/publications/paper/ps5t}

\bibitem{DBLP:conf/apn/EsparzaLS10}
Esparza, J., Leucker, M., Schlund, M.: Learning workflow petri nets. In:
  Lilius, J., Penczek, W. (eds.) Applications and Theory of Petri Nets, 31st
  International Conference, {PETRI} {NETS} 2010, Braga, Portugal, June 21-25,
  2010. Proceedings. Lecture Notes in Computer Science, vol.~6128, pp.
  206--225. Springer (2010). \doi{10.1007/978-3-642-13675-7\_13},
  \url{https://doi.org/10.1007/978-3-642-13675-7\_13}

\bibitem{DBLP:conf/rv/FainekosH019}
Fainekos, G., Hoxha, B., Sankaranarayanan, S.: Robustness of specifications and
  its applications to falsification, parameter mining, and runtime monitoring
  with s-taliro. In: Finkbeiner, B., Mariani, L. (eds.) Runtime Verification -
  19th International Conference, {RV} 2019, Porto, Portugal, October 8-11,
  2019, Proceedings. Lecture Notes in Computer Science, vol. 11757, pp. 27--47.
  Springer (2019). \doi{10.1007/978-3-030-32079-9\_3},
  \url{https://doi.org/10.1007/978-3-030-32079-9\_3}

\bibitem{DBLP:journals/tcs/FainekosP09}
Fainekos, G.E., Pappas, G.J.: Robustness of temporal logic specifications for
  continuous-time signals. Theor. Comput. Sci.  \textbf{410}(42),  4262--4291
  (2009). \doi{10.1016/j.tcs.2009.06.021},
  \url{https://doi.org/10.1016/j.tcs.2009.06.021}

\bibitem{DBLP:journals/tse/FujiwaraBKAG91}
Fujiwara, S., von Bochmann, G., Khendek, F., Amalou, M., Ghedamsi, A.: Test
  selection based on finite state models. {IEEE} Trans. Software Eng.
  \textbf{17}(6),  591--603 (1991). \doi{10.1109/32.87284},
  \url{https://doi.org/10.1109/32.87284}

\bibitem{DBLP:journals/ngc/Hasuo17}
Hasuo, I.: Metamathematics for systems design - comprehensive transfer of
  formal methods techniques to cyber-physical systems. New Gener. Comput.
  \textbf{35}(3),  271--305 (2017). \doi{10.1007/s00354-017-0023-1},
  \url{https://doi.org/10.1007/s00354-017-0023-1}

\bibitem{DBLP:conf/se/HerberAL21}
Herber, P., Adelt, J., Liebrenz, T.: Formal verification of intelligent
  cyber-physical systems with the interactive theorem prover keymaera {X}. In:
  G{\"{o}}tz, S., Linsbauer, L., Schaefer, I., Wortmann, A. (eds.) Proceedings
  of the Software Engineering 2021 Satellite Events, Braunschweig/Virtual,
  Germany, February 22 - 26, 2021. {CEUR} Workshop Proceedings, vol.~2814.
  CEUR-WS.org (2021), \url{http://ceur-ws.org/Vol-2814/short-A3-2.pdf}

\bibitem{DBLP:conf/dagstuhl/HowarS16}
Howar, F., Steffen, B.: Active automata learning in practice - an annotated
  bibliography of the years 2011 to 2016. In: Bennaceur, A., H{\"{a}}hnle, R.,
  Meinke, K. (eds.) Machine Learning for Dynamic Software Analysis: Potentials
  and Limits - International Dagstuhl Seminar 16172, Dagstuhl Castle, Germany,
  April 24-27, 2016, Revised Papers. Lecture Notes in Computer Science, vol.
  11026, pp. 123--148. Springer (2018). \doi{10.1007/978-3-319-96562-8\_5},
  \url{https://doi.org/10.1007/978-3-319-96562-8\_5}

\bibitem{DBLP:conf/cpsweek/HoxhaAF14}
Hoxha, B., Abbas, H., Fainekos, G.E.: Benchmarks for temporal logic
  requirements for automotive systems. In: Frehse, G., Althoff, M. (eds.) 1st
  and 2nd International Workshop on Applied veRification for Continuous and
  Hybrid Systems, ARCH@CPSWeek 2014, Berlin, Germany, April 14, 2014 /
  ARCH@CPSWeek 2015, Seattle, WA, USA, April 13, 2015. EPiC Series in
  Computing, vol.~34, pp. 25--30. EasyChair (2014),
  \url{https://easychair.org/publications/paper/4bfq}

\bibitem{DBLP:conf/cpsweek/HoxhaAF14a}
Hoxha, B., Abbas, H., Fainekos, G.E.: Using s-taliro on industrial size
  auimmlertomotive models. In: Frehse, G., Althoff, M. (eds.) 1st and 2nd
  International Workshop on Applied veRification for Continuous and Hybrid
  Systems, ARCH@CPSWeek 2014, Berlin, Germany, April 14, 2014 / ARCH@CPSWeek
  2015, Seattle, WA, USA, April 13, 2015. EPiC Series in Computing, vol.~34,
  pp. 113--119. EasyChair (2014),
  \url{https://easychair.org/publications/paper/r8gZ}

\bibitem{DBLP:conf/rv/IsbernerHS14}
Isberner, M., Howar, F., Steffen, B.: The {TTT} algorithm: {A} redundancy-free
  approach to active automata learning. In: Bonakdarpour, B., Smolka, S.A.
  (eds.) Runtime Verification - 5th International Conference, {RV} 2014,
  Toronto, ON, Canada, September 22-25, 2014. Proceedings. Lecture Notes in
  Computer Science, vol.~8734, pp. 307--322. Springer (2014).
  \doi{10.1007/978-3-319-11164-3\_26},
  \url{https://doi.org/10.1007/978-3-319-11164-3\_26}

\bibitem{DBLP:conf/cav/IsbernerHS15}
Isberner, M., Howar, F., Steffen, B.: The open-source learnlib - {A} framework
  for active automata learning. In: Kroening, D., Pasareanu, C.S. (eds.)
  Computer Aided Verification - 27th International Conference, {CAV} 2015, San
  Francisco, CA, USA, July 18-24, 2015, Proceedings, Part {I}. Lecture Notes in
  Computer Science, vol.~9206, pp. 487--495. Springer (2015).
  \doi{10.1007/978-3-319-21690-4\_32},
  \url{https://doi.org/10.1007/978-3-319-21690-4\_32}

\bibitem{DBLP:conf/kbse/KhosrowjerdiM18}
Khosrowjerdi, H., Meinke, K.: Learning-based testing for autonomous systems
  using spatial and temporal requirements. In: Perrouin, G., Acher, M., Cordy,
  M., Devroey, X. (eds.) Proceedings of the 1st International Workshop on
  Machine Learning and Software Engineering in Symbiosis, MASES@ASE 2018,
  Montpellier, France, September 3, 2018. pp. 6--15. {ACM} (2018).
  \doi{10.1145/3243127.3243129}, \url{https://doi.org/10.1145/3243127.3243129}

\bibitem{kirkpatrick1983optimization}
Kirkpatrick, S., Gelatt, C.D., Vecchi, M.P.: Optimization by simulated
  annealing. science  \textbf{220}(4598),  671--680 (1983)

\bibitem{DBLP:conf/fm/LinH14}
Lin, S., Hsiung, P.: Compositional synthesis of concurrent systems through
  causal model checking and learning. In: Jones, C.B., Pihlajasaari, P., Sun,
  J. (eds.) {FM} 2014: Formal Methods - 19th International Symposium,
  Singapore, May 12-16, 2014. Proceedings. Lecture Notes in Computer Science,
  vol.~8442, pp. 416--431. Springer (2014).
  \doi{10.1007/978-3-319-06410-9\_29},
  \url{https://doi.org/10.1007/978-3-319-06410-9\_29}

\bibitem{DBLP:conf/formats/MalerN04}
Maler, O., Nickovic, D.: Monitoring temporal properties of continuous signals.
  In: Lakhnech, Y., Yovine, S. (eds.) Formal Techniques, Modelling and Analysis
  of Timed and Fault-Tolerant Systems, Joint International Conferences on
  Formal Modelling and Analysis of Timed Systems, {FORMATS} 2004 and Formal
  Techniques in Real-Time and Fault-Tolerant Systems, {FTRTFT} 2004, Grenoble,
  France, September 22-24, 2004, Proceedings. Lecture Notes in Computer
  Science, vol.~3253, pp. 152--166. Springer (2004).
  \doi{10.1007/978-3-540-30206-3\_12},
  \url{https://doi.org/10.1007/978-3-540-30206-3\_12}

\bibitem{DBLP:journals/isse/MeijerP19}
Meijer, J., van~de Pol, J.: Sound black-box checking in the learnlib. Innov.
  Syst. Softw. Eng.  \textbf{15}(3-4),  267--287 (2019).
  \doi{10.1007/s11334-019-00342-6},
  \url{https://doi.org/10.1007/s11334-019-00342-6}

\bibitem{DBLP:conf/pts/MeinkeN10}
Meinke, K., Niu, F.: A learning-based approach to unit testing of numerical
  software. In: Petrenko, A., da~Silva~Sim{\~{a}}o, A., Maldonado, J.C. (eds.)
  Testing Software and Systems - 22nd {IFIP} {WG} 6.1 International Conference,
  {ICTSS} 2010, Natal, Brazil, November 8-10, 2010. Proceedings. Lecture Notes
  in Computer Science, vol.~6435, pp. 221--235. Springer (2010).
  \doi{10.1007/978-3-642-16573-3\_16},
  \url{https://doi.org/10.1007/978-3-642-16573-3\_16}

\bibitem{DBLP:conf/sefm/MeinkeN15}
Meinke, K., Nycander, P.: Learning-based testing of distributed microservice
  architectures: Correctness and fault injection. In: Bianculli, D., Calinescu,
  R., Rumpe, B. (eds.) Software Engineering and Formal Methods - {SEFM} 2015
  Collocated Workshops: ATSE, HOFM, MoKMaSD, and VERY*SCART, York, UK,
  September 7-8, 2015, Revised Selected Papers. Lecture Notes in Computer
  Science, vol.~9509, pp. 3--10. Springer (2015).
  \doi{10.1007/978-3-662-49224-6\_1},
  \url{https://doi.org/10.1007/978-3-662-49224-6\_1}

\bibitem{DBLP:conf/icst/MeinkeS13}
Meinke, K., Sindhu, M.A.: Lbtest: {A} learning-based testing tool for reactive
  systems. In: Sixth {IEEE} International Conference on Software Testing,
  Verification and Validation, {ICST} 2013, Luxembourg, Luxembourg, March
  18-22, 2013. pp. 447--454. {IEEE} Computer Society (2013).
  \doi{10.1109/ICST.2013.62}, \url{https://doi.org/10.1109/ICST.2013.62}

\bibitem{DBLP:conf/sigsoft/2015}
Nitto, E.D., Harman, M., Heymans, P. (eds.): Proceedings of the 2015 10th Joint
  Meeting on Foundations of Software Engineering, {ESEC/FSE} 2015, Bergamo,
  Italy, August 30 - September 4, 2015. {ACM} (2015). \doi{10.1145/2786805},
  \url{https://doi.org/10.1145/2786805}

\bibitem{DBLP:conf/forte/PeledVY99}
Peled, D.A., Vardi, M.Y., Yannakakis, M.: Black box checking. In: Wu, J.,
  Chanson, S.T., Gao, Q. (eds.) Formal Methods for Protocol Engineering and
  Distributed Systems, {FORTE} {XII} / {PSTV} XIX'99, {IFIP} {TC6} {WG6.1}
  Joint International Conference on Formal Description Techniques for
  Distributed Systems and Communication Protocols {(FORTE} {XII)} and Protocol
  Specification, Testing and Verification {(PSTV} XIX), October 5-8, 1999,
  Beijing, China. {IFIP} Conference Proceedings, vol.~156, pp. 225--240. Kluwer
  (1999)

\bibitem{DBLP:conf/focs/Pnueli77}
Pnueli, A.: The temporal logic of programs. In: 18th Annual Symposium on
  Foundations of Computer Science, Providence, Rhode Island, USA, 31 October -
  1 November 1977. pp. 46--57. {IEEE} Computer Society (1977).
  \doi{10.1109/SFCS.1977.32}, \url{https://doi.org/10.1109/SFCS.1977.32}

\bibitem{DBLP:journals/corr/abs-2012-00319}
Sato, S., Waga, M., Hasuo, I.: Constrained optimization for falsification and
  conjunctive synthesis. CoRR  \textbf{abs/2012.00319} (2020),
  \url{https://arxiv.org/abs/2012.00319}

\bibitem{DBLP:conf/sfm/SteffenHM11}
Steffen, B., Howar, F., Merten, M.: Introduction to active automata learning
  from a practical perspective. In: Bernardo, M., Issarny, V. (eds.) Formal
  Methods for Eternal Networked Software Systems - 11th International School on
  Formal Methods for the Design of Computer, Communication and Software
  Systems, {SFM} 2011, Bertinoro, Italy, June 13-18, 2011. Advanced Lectures.
  Lecture Notes in Computer Science, vol.~6659, pp. 256--296. Springer (2011).
  \doi{10.1007/978-3-642-21455-4\_8},
  \url{https://doi.org/10.1007/978-3-642-21455-4\_8}

\bibitem{DBLP:conf/csl/TabuadaN16}
Tabuada, P., Neider, D.: Robust linear temporal logic. In: Talbot, J., Regnier,
  L. (eds.) 25th {EACSL} Annual Conference on Computer Science Logic, {CSL}
  2016, August 29 - September 1, 2016, Marseille, France. LIPIcs, vol.~62, pp.
  10:1--10:21. Schloss Dagstuhl - Leibniz-Zentrum f{\"{u}}r Informatik (2016).
  \doi{10.4230/LIPIcs.CSL.2016.10},
  \url{https://doi.org/10.4230/LIPIcs.CSL.2016.10}

\bibitem{DBLP:conf/hybrid/Waga20}
Waga, M.: Falsification of cyber-physical systems with robustness-guided
  black-box checking. In: Ames, A.D., Seshia, S.A., Deshmukh, J. (eds.) {HSCC}
  '20: 23rd {ACM} International Conference on Hybrid Systems: Computation and
  Control, Sydney, New South Wales, Australia, April 21-24, 2020. pp.
  11:1--11:13. {ACM} (2020). \doi{10.1145/3365365.3382193},
  \url{https://doi.org/10.1145/3365365.3382193}

\bibitem{DBLP:conf/fmcad/YamaguchiKDS16}
Yamaguchi, T., Kaga, T., Donz{\'{e}}, A., Seshia, S.A.: Combining requirement
  mining, software model checking and simulation-based verification for
  industrial automotive systems. In: Piskac, R., Talupur, M. (eds.) 2016 Formal
  Methods in Computer-Aided Design, {FMCAD} 2016, Mountain View, CA, USA,
  October 3-6, 2016. pp. 201--204. {IEEE} (2016).
  \doi{10.1109/FMCAD.2016.7886680},
  \url{https://doi.org/10.1109/FMCAD.2016.7886680}

\bibitem{DBLP:journals/tcad/ZhangESAH18}
Zhang, Z., Ernst, G., Sedwards, S., Arcaini, P., Hasuo, I.: Two-layered
  falsification of hybrid systems guided by monte carlo tree search. {IEEE}
  Trans. Comput. Aided Des. Integr. Circuits Syst.  \textbf{37}(11),
  2894--2905 (2018). \doi{10.1109/TCAD.2018.2858463},
  \url{https://doi.org/10.1109/TCAD.2018.2858463}

\end{thebibliography}

\LongVersion{\appendix
\section{Proof of \cref{th:rewrite-correct}}\label{apdx:proof:rewrite-correct}

In the proof of \cref{th:rewrite-correct}, we use the following notation.

\begin{definition}
 [$\varphi \succeq \varphi'$]
 \label{def:ltl-strength}
 For LTL formulas $\varphi$ and $\varphi'$, $\varphi'$ is \emph{stronger} than $\varphi$
 if for any $\pi \in (\mathcal{P}(\AP))^\omega$ and $k \in \mathbb{N}$, $(\pi, k) \models \varphi'$ implies $(\pi, k) \models \varphi$.
 For such $\varphi$ and $\varphi'$, we denote $\varphi \succeq \varphi'$.
\end{definition}


The following proves \cref{th:rewrite-correct}.

\begin{proof}
  We prove \cref{th:rewrite-correct} by induction on the structure 
of $(\varphi, \psi) \in {\rightarrowtail}$.
  \begin{enumerate}
    \item When $\exists \mu, \nu \in \LTL. \ \varphi = \mu \lor \nu \ \mbox{and} \ \psi = \mu \land \nu$.
      We choose arbitrary $\pi \in (\mathcal{P}(\AP))^\omega$ and $k \in \mathbb{N}$.
      We assume $(\pi, k) \models \mu \land \nu$.
      By the definition of the semantics of LTL formulas in \cref{definition:LTLsemantics},
      we have $(\pi, k) \models \mu \ \mbox{and} \ (\pi, k) \models \nu$.
      Therefore, we have $(\pi, k) \models \mu \ \mbox{or} \ (\pi, k) \models \nu$.
      By \cref{definition:LTLsemantics}, we have $(\pi, k) \models \mu \lor \nu$.
      We thus get $\mu \lor \nu \succeq \mu \land \nu$. This is $\varphi \succeq \psi$.
    \item When $\exists \mu \in \LTL. \ \varphi = \Evt \mu \ \mbox{and} \ \psi = \Glb \Evt \mu$.\label{proof:rewrite-correct:Evt}
      We choose arbitrary $\pi \in (\mathcal{P}(\AP))^\omega$ and $k \in \mathbb{N}$.
      We assume $(\pi, k) \models \Glb \Evt \mu$.
      Expanding the syntactic abbreviations of LTL formulas, we have $(\pi, k) \models \neg (\top \Until (\neg (\top \Until \mu)))$.
      By \cref{definition:LTLsemantics}, we have $(\pi, k) \nvDash \top \Until (\neg (\top \Until \mu))$,
      and it follows that $\forall l \in [k, \infty). \ (\pi, l) \nvDash \neg (\top \Until \mu) \ \lor \ \exists m \in \{k, k+1, \cdots, l\}. \ (\pi, m) \nvDash \top$.
      Here, since $(\pi, n) \nvDash \top$ does not hold for any natural number $n$,
      we have $(\pi, k) \nvDash \neg (\top \Until \mu)$.
      By \cref{definition:LTLsemantics}, we have $(\pi, k) \models \top \Until \mu$.
      Using the definition of the notation of $\Evt$ operator, we have $(\pi, k) \models \Evt \mu$.
      We thus get $\Evt \mu \succeq \Glb \Evt \mu$. This is $\varphi \succeq \psi$.
    \item When $\exists \mu \in \LTL. \ \varphi = \Glb \Evt \mu \ \mbox{and} \ \psi = \Evt \Glb \mu$.\label{proof:rewrite-correct:GlbEvt}
      We choose arbitrary $\pi \in (\mathcal{P}(\AP))^\omega$ and $k \in \mathbb{N}$.
      We assume $(\pi, k) \models \Evt \Glb \mu$.
      Expanding the syntactic abbreviations of LTL formulas, we have $(\pi, k) \models \top \Until (\neg (\top \Until \neg \mu))$.
      By \cref{definition:LTLsemantics}, we have $\exists l \in [k, \infty). \ (\pi, l) \models \neg (\top \Until \neg \mu) \ \ \land \ \forall m \in \{k, k+1, \cdots, l\}. \ (\pi, m) \models \top$.
      From $(\pi, l) \models \neg (\top \Until \neg \mu)$, it follows that $(\pi, l) \nvDash \top \Until \neg \mu$,
      and we have $\forall l' \in [l, \infty). \ (\pi, l') \nvDash \neg \mu \ \ \lor \exists m' \in \{l, l+1, \cdots, l'\}. \ (\pi, m') \nvDash \top$.
      Here, since $(\pi, n) \nvDash \top$ does not hold for any natural number $n$,
      we have $\forall l' \in [l, \infty). \ (\pi, l') \nvDash \neg \mu$.
      By \cref{definition:LTLsemantics}, we have $\forall l' \in [l, \infty). \ (\pi, l') \models \mu$.
      In other words, there exists a natural number $l \in [k, \infty)$,
      and for any natural number $l'$ after $l$, we have $(\pi, l') \models \mu$.
      Therefore, we have $\forall p \in [k, \infty). \ \exists q \in [p, \infty). \ (\pi, q) \models \mu$.
      Since $(\pi, n) \models \top$ holds for any natural number $n$, we have $\forall p \in [k, \infty). \ \exists q \in [p, \infty). \ (\pi, q) \models \mu \ \ \land \forall r \in \{p, p+1, \cdots, q\}. \ (\pi, r) \models \top$.
      By the definition of $\Until$ operator in \cref{definition:LTLsemantics},
      we have $\forall p \in [k, \infty). \ (\pi, p) \models \top \Until \mu$.
      Furthermore, we have $\forall p \in [k, \infty). \ (\pi, p) \models \top \Until \mu \ \ \lor \exists r' \in \{k, k+1, \cdots, p\}. \ (\pi, r') \nvDash \top$.
      We take the whole negative and use the definition of $\Until$ operator in \cref{definition:LTLsemantics},
      then we have $(\pi, k) \nvDash \top \Until (\neg (\top \Until \mu))$.
      By the definition of $\neg$ operator in \cref{definition:LTLsemantics} and the definition of the syntactic abbreviations of LTL formulas, we have $(\pi, k) \models \Glb \Evt \mu$.
      We thus get $\Glb \Evt \mu \succeq \Evt \Glb \mu$. This is $\varphi \succeq \psi$.
    \item When $\exists \mu \in \LTL. \ \varphi = \Evt \Glb \mu \ \mbox{and} \ \psi = \Glb \mu$.
      We choose arbitrary $\pi \in (\mathcal{P}(\AP))^\omega$ and $k \in \mathbb{N}$.
      We assume $(\pi, k) \models \Glb \mu$.
      Expanding the syntactic abbreviations of LTL formulas, we have $(\pi, k) \models \neg (\top \Until \neg \mu)$.
      By \cref{definition:LTLsemantics}, we have $(\pi, k) \nvDash \top \Until \neg \mu$.
      Furthermore, we have $\forall l \in [k, \infty). \ (\pi, l) \nvDash \neg \mu \ \ \lor \ \exists m \in \{k, k+1, \cdots l\}. \ (\pi, k)  \nvDash \top$.
      Here, since $(\pi, n) \nvDash \top$ does not hold for any natural number $n$,
      we have $\forall l \in [k, \infty). \ (\pi, l) \nvDash \neg \mu$.
      By \cref{definition:LTLsemantics}, we have $\forall l \in [k, \infty). \ (\pi, l) \models \mu$.
      Since we have $k \in [k, \infty)$ and $\forall l \in [k, \infty). \ (\pi, l) \models \mu$, we have $\exists l \in [k, \infty). \ \Big( \forall l' \in [l, \infty). \ (\pi, l') \models \mu \ \ \lor \exists m' \in \{l, l+1, \cdots, l'\}. \ (\pi, m') \nvDash \top \Big)\ \ \land \ \forall m \in \{k, k+1, \cdots, l\}. \ (\pi, m) \models \top$.
      By \cref{definition:LTLsemantics}, we have $(\pi, k) \models \Evt \Glb \mu$.
      We thus get $\Evt \Glb \mu \succeq \Glb \mu$. This is $\varphi \succeq \psi$.
    \item When $\exists \mu \in \LTL. \ \exists i, j \in \mathbb{N} \cup \{\infty\}. \ \varphi = \Evt_{[i, j)} \mu \ \mbox{and} \ \psi = \Glb_{[i, j)} \mu$.
      We choose arbitrary $\pi \in (\mathcal{P}(\AP))^\omega$ and $k \in \mathbb{N}$.
      We assume $(\pi, k) \models \Glb_{[i, j)} \mu$.
      Expanding the syntactic abbreviations of LTL formulas, we have $(\pi, k) \models \neg (\top \Until[[i, j)] \neg \mu)$.
      By \cref{definition:LTLsemantics}, we have $\forall l \in [k+i, k+j). \ (\pi, l) \models \mu \ \ \lor \ \exists m \in \{k, k+1, \cdots, l\}. \ (\pi, m) \nvDash \top$.
      Here, since $(\pi, n) \nvDash \top$ does not hold for any natural number $n$,
      we have $\forall l \in [k+i, k+j). \ (\pi, l) \models \mu$.
      Therefore, we have $\exists l' \in [k+i, k+j). \ (\pi, l') \models \mu$.
      Since $(\pi, n) \models \top$ holds for any natural number $n$,
      we have $\exists l' \in [k+i, k+j). \ (\pi, l') \models \mu \ \ \land \ \forall m' \in \{k, k+1, \cdots, l'\}. \ (\pi, m') \models \top$.
      By \cref{definition:LTLsemantics}, we have $(\pi, k) \models \top \Until_{[i, j)} \mu$.
      Using the notation of LTL formulas, we have $(\pi, k) \models \Evt_{[i, j)} \mu$.
      We thus get $\Evt_{[i, j)} \mu \succeq \Glb_{[i, j)} \mu$. This is $\varphi \succeq \psi$.
    \item When $\exists \mu, \nu \in \LTL. \ \varphi = \mu \Until \nu \ \mbox{and} \ \psi = \Glb \mu \land \Glb \Evt \nu$.
      We choose arbitrary $\pi \in (\mathcal{P}(\AP))^\omega$ and $k \in \mathbb{N}$.
      We assume $(\pi, k) \models \Glb \mu \land \Glb \Evt \nu$.
      By the definition of the semantic of LTL formulas~\cref{definition:LTLsemantics},
      we have $(\pi, k) \models \Glb \mu$ and $(\pi, k) \models \Glb \Evt \nu$.
      Expanding the syntactic abbreviations of LTL formulas, from $(\pi, k) \models \Glb \mu$, it follows that $\forall l \in [k, \infty). \ (\pi, l) \models \mu \ \lor \exists m \in \{k, k+1, \cdots, l\}. \ (\pi, m) \nvDash \top$.
      Here, since $(\pi, n) \nvDash \top$ does not hold for any natural number $n$,
      we have $\forall l \in [k, \infty). \ (\pi, l) \models \mu$.
      Also, from $(\pi, k) \models \Glb \Evt \nu$, doing the same as \ref{proof:rewrite-correct:Evt}.,
      we have $\forall l \in [k, \infty). \ (\pi, l) \models (\top \Until \nu) \ \lor \ \exists m \in \{k, k+1, \cdots, l\}. \ (\pi, m) \nvDash \top$.
      Since $(\pi, n) \nvDash \top$ does not hold for any natural number $n$,
      we have $\forall l \in [k, \infty). \ (\pi, l) \models (\top \Until \nu)$.
      Since $k \in [k, \infty)$, we have $(\pi, k) \models \top \Until \nu$.
      By \cref{definition:LTLsemantics}, we have $\exists l' \in [k, \infty). (\pi, l') \models \nu \land \forall m' \in \{k, k+1, \cdots, l'\}. \ (\pi, m') \models \top$.
      Therefore, from $\exists l' \in [k, \infty). (\pi, l') \models \nu$ and $\forall l \in [k, \infty). \ (\pi, l) \models \mu$,
      it follows that $\exists l' \in [k, \infty). (\pi, l') \models \nu \ \ \land \forall r \in \{k, k+1, \cdots l'\}. \ (\pi, r) \models \mu$.
      By \cref{definition:LTLsemantics}, we have $(\pi, k) \models \mu \Until \nu$
      We thus get $\Glb \mu \land \Glb \Evt \nu \succeq \mu \Until \nu$.
      This is $\varphi \succeq \psi$.
    \item When $\exists \mu \in \LTL$. $\exists i, j, i', j' \in \mathbb{N}\cup\{\infty\}$. $[i, j) \supsetneq [i', j')$ and $\varphi = \Evt_{[i, j)} \mu$ and $\psi = \Evt_{[i', j')} \mu$.
      We choose arbitrary $\pi \in (\mathcal{P}(\AP))^\omega$ and $k \in \mathbb{N}$.
      We assume $(\pi, k) \models \Evt_{[i', j')} \mu$.
      Expanding the syntactic abbreviations of LTL, we have $(\pi, k) \models \top \Until[[i', j')] \mu$.
      By the semantics of LTL formula~\cref{definition:LTLsemantics}, there exists $l \in [k + i', k + j')$ such that $(\pi, l) \models \mu$ and $\forall m \in k, k + 1, \dots, l. \ (\pi, m) \models \top$.
      Since $[i, j) \supsetneq [i', j')$, we have $l \in [i, j)$.
      Since $(\pi, n) \models \top$ holds for any natural number $n$, we have $(\pi, l') \models \mu$ and $\forall m \in k, k + 1, \dots, l. \ (\pi, m) \models \top$.
      By \cref{definition:LTLsemantics}, we have $(\pi, k) \models \top \Until[[i, j)] \mu$.
      By the syntactic abbreviations, we have $(\pi, k) \models \Evt_{[i, j)} \mu$.
      We thus get $\Evt_{[i, j)} \mu \succeq \Evt_{[i', j')} \mu$.
      This is $\varphi \succeq \psi$.
    \item When $\exists \mu, \nu \in \LTL. \ \nu \rightarrowtail \mu \ \mbox{and} \ \varphi = \neg \mu \ \mbox{and} \ \psi = \neg \nu$.
      By induction hypothesis, we have $\nu \succeq \mu$.
      We choose arbitrary $\pi \in (\mathcal{P}(\AP))^\omega$ and $k \in \mathbb{N}$.
      We assume $(\pi, k) \models \neg \nu$.
      By the semantics of LTL formula~\cref{definition:LTLsemantics}, we have $(\pi, k) \not\models \nu$.
      From $\nu \succeq \mu$, it follows that $(\pi, k) \models \mu \implies (\pi, k) \models \nu$.
      Taking the contrapositive, we have $(\pi, k) \not\models \nu \implies (\pi, k) \not\models \mu$.
      Therefore, we have $(\pi, k) \not\models \mu$.
      By \cref{definition:LTLsemantics}, we have $(\pi, k) \models \neg \mu$.
      By \cref{def:ltl-strength}, we have $\neg \mu \succeq \neg \nu$.
      This is $\varphi \succeq \psi$.
    \item When $\exists \mu, \mu', \nu \in \LTL$. $\mu \rightarrowtail \mu'$ and $\varphi = \mu \lor \nu$ and $\psi = \mu' \lor \nu$.
      By induction hypothesis, we have $\mu \succeq \mu'$.
      We choose arbitrary $\pi \in (\mathcal{P}(\AP))^\omega$ and $k \in \mathbb{N}$.
      We assume $(\pi, k) \models \mu' \lor \nu$.
      By the semantics of LTL formula~\cref{definition:LTLsemantics}, we have $(\pi, k) \models \mu'$ or $(\pi, k) \models \nu$.
      From $\mu \succeq \mu'$, it follows that $(\pi, k) \models \mu' \implies (\pi, k) \models \mu$.
      Therefore, we have $(\pi, k) \models \mu$ or $(\pi, k) \models \nu$.
      By \cref{definition:LTLsemantics}, we have $(\pi, k) \models \mu \lor \nu$.
      By \cref{def:ltl-strength}, we have $\mu \lor \nu \succeq \mu' \lor \nu$.
      This is $\varphi \succeq \psi$.
    \item When $\exists \mu, \nu, \nu' \in \LTL$. $\nu \rightarrowtail \nu'$ and $\varphi = \mu \lor \nu$ and $\psi = \mu \lor \nu'$.
      By induction hypothesis, we have $\nu \succeq \nu'$.
      We choose arbitrary $\pi \in (\mathcal{P}(\AP))^\omega$ and $k \in \mathbb{N}$.
      We assume $(\pi, k) \models \mu \lor \nu'$.
      By the semantics of LTL formula~\cref{definition:LTLsemantics}, we have $(\pi, k) \models \mu$ or $(\pi, k) \models \nu'$.
      From $\nu \succeq \nu'$, it follows that $(\pi, k) \models \nu' \implies (\pi, k) \models \nu$.
      Therefore, we have $(\pi, k) \models \mu$ or $(\pi, k) \models \nu$.
      By \cref{definition:LTLsemantics}, we have $(\pi, k) \models \mu \lor \nu$.
      By \cref{def:ltl-strength}, we have $\mu \lor \nu \succeq \mu \lor \nu'$.
      This is $\varphi \succeq \psi$.
    \item When $\exists \mu, \nu \in \LTL$. $\mu \rightarrowtail \nu$ and $\varphi = \Next \mu$ and $\psi = \Next \nu$.
      By induction hypothesis, we have $\mu \succeq \nu$.
      We choose arbitrary $\pi \in (\mathcal{P}(\AP))^\omega$ and $k \in \mathbb{N}$.
      We assume $(\pi, k) \models \Next \nu$.
      By the semantics of LTL formula~\cref{definition:LTLsemantics}, we have $(\pi, k + 1) \models \nu$.
      From $\mu \succeq \nu$, it follows that $(\pi, k + 1) \models \nu \implies (\pi, k + 1) \models \mu$.
      Therefore, we have $(\pi, k + 1) \models \mu$.
      By \cref{definition:LTLsemantics}, we have $(\pi, k) \models \Next \mu$.
      By \cref{def:ltl-strength}, we have $\Next \mu \succeq \Next \nu$.
      This is $\varphi \succeq \psi$.
    \item When $\exists \mu, \nu, \nu' \in \LTL$. $\exists i, j \in \mathbb{N}\cup\{\infty\}$. $\nu \rightarrowtail \nu'$ and $\varphi = \mu \Until[[i, j)] \nu$ and $\psi = \mu \ \Until[[i, j)] \nu'$.
      By induction hypothesis, we have $\nu \succeq \nu'$.
      We choose arbitrary $\pi \in (\mathcal{P}(\AP))^\omega$ and $k \in \mathbb{N}$.
      We assume $(\pi, k) \models \mu \Until[[i, j)]\nu'$.
      By the semantics of LTL formula~\cref{definition:LTLsemantics}, there exists $l \in [i + k, j + k)$ such that $(\pi, l) \models \nu'$ and $\forall m \in k, k + 1, \dots, l. \ (\pi, m) \models \mu$.
      From $\nu \succeq \nu'$, it follows that $(\pi, k) \models \nu' \implies (\pi, k) \models \nu$.
      Therefore, we have $(\pi, l) \models \nu$ and $\forall m \in k, k + 1, \dots, l. \ (\pi, m) \models \mu$.
      By \cref{definition:LTLsemantics}, we have $(\pi, k) \models \mu \Until[[i, j)] \nu$.
      By \cref{def:ltl-strength}, we have $\mu \Until[[i, j)] \nu \succeq \mu \Until[[i, j)] \nu'$.
      This is $\varphi \succeq \psi$.
    \item When $\exists \mu \in \LTL$. $\varphi \rightarrowtail \mu$ and $\mu \rightarrowtail \psi$.
      By induction hypothesis, we have $\varphi \succeq \mu$ and $\mu \succeq \psi$.
      We choose arbitrary $\pi \in (\mathcal{P}(\AP))^\omega$ and $k \in \mathbb{N}$.
      We assume $(\pi, k) \models \psi$.
      By $\mu \succeq \psi$, we have $(\pi, k) \models \mu$.
      By $\varphi \succeq \mu$, we have $(\pi, k) \models \varphi$.
      By \cref{def:ltl-strength}, we have $\varphi \succeq \psi$.
  \end{enumerate}
 \qed{}
\end{proof}



}

\end{document}